\documentstyle[epsf,graphicx,color]{aa}

\newcommand{\AaAS}{A\&As}
\newcommand{\AaA}{A\&A}

\newcommand{\ApJ}{ApJ}
\newcommand{\MNRAS}{MNRAS}

\newcommand{\PhRvD}{Phys. Rev. D}
\newcommand{\PhRvL}{Phys. Rev. Lett.}

\newcommand{\be}{\begin{equation}}
\newcommand{\ee}{\end{equation}}
\newcommand{\bea}{\begin{eqnarray}}
\newcommand{\eea}{\end{eqnarray}}
\newcommand{\bk}{{\bf k}}
\newcommand{\bq}{{\bf q}}

\newcommand{\bx}{{\bf x}}
\newcommand{\ba}{{\bf a}}
\newcommand{\dk}{d^2\!k\,}
\newcommand{\dx}{d^2\!x\,}
\definecolor{light}{gray}{.75}
\begin{document}
\title{Non-Gaussianity: Comparing wavelet and Fourier based methods} 
\author{N. Aghanim\inst{1} \and M. Kunz\inst{2,}\inst{3} \and P.G. Castro\inst{2} \and O. Forni\inst{1}}
\offprints{}
\institute{IAS-CNRS, B\^atiment 121, Universit\'e Paris Sud, F-91405 Orsay, 
France \and Denys Wilkinson building, Oxford University, OX1 3RH, Oxford, U.K. 
\and Astronomy Centre, University of Sussex, BN1 9QJ, Brighton, U.K.}
\date{Received date / accepted date}
\abstract{In the context of the present and future Cosmic 
Microwave Background (CMB) experiments, 
going beyond the information provided by the power spectrum
has become necessary in order to tightly constrain the
cosmological model. The non-Gaussian signatures in the CMB
represent a very promising tool to probe the early
universe and the structure formation epoch.
We present the results of a comparison between two families 
of non-Gaussian estimators: The first act on the wavelet space
(skewness and excess kurtosis of the wavelet coefficients) and the 
second group on the Fourier space (bi- and trispectrum). 
We compare the relative sensitivities of these estimators by
applying them to three different data sets meant to reproduce 
the majority of possible non-Gaussian contributions to the CMB.
We find that the skewness in the wavelet space is slightly
more sensitive than the bispectrum. For the four point estimators, 
we find that the excess kurtosis of the wavelet coefficients has 
very similar capabilities than the diagonal trispectrum while a
near-diagonal trispectrum seems to be less sensitive to 
non-Gaussian signatures. 
}
\maketitle
\markboth{Non-Gaussianity: Comparing wavelet and Fourier based methods}{}

\section{Introduction}
Over the last few years, the amount of available data on the
Cosmic Microwave Background (CMB) has increased dramatically.
Recent measurements of the CMB power spectrum (BOOMERANG, MAXIMA, 
DASI, CBI, ARCHEOPS, VSA, ACBAR...) have started
to shed light on the origin and large scale structure of our universe.
In particular, the determination of the density parameter 
$\Omega\simeq 1$ has been obtained from the position of the first 
acoustic peak.
Together with results from other experiments, which probe e.g.
the distribution of galaxies and the distances to type-Ia supernovae, 
a coherent image of todays universe has started to emerge (see e.g. 
\cite{Wang2002}). Testing the detailed statistical nature of the 
CMB anisotropies has become 
not only one of the major goals of CMB cosmology but it is now also
within our reach. The MAP satellite (and also the future
Planck mission) will provide us with an all sky survey of the CMB with 
good angular resolution that can be used to search for non-Gaussian
signatures. 

A Gaussian random field is completely determined
by its power spectrum. But many models of the early universe,
like inflation (e.g. \cite{bartolo2002}, \cite{bernardeau2002a} and 
references therein), super strings or topological defects, predict 
non-Gaussian contributions to the
initial fluctuations \cite[]{luo94a,jaffe94,gangui94}. 
Furthermore, any non-linearity in their evolution introduces additional
non-Gaussian signatures. An extreme example is gravitational clustering,
which makes it quite difficult to extract any primordial
non-Gaussianity from the galaxy distribution \cite[]{durrer2000}. 
The CMB on the other hand is to first order free of this complication
and is therefore ideally suited to study the early universe. 
Nevertheless, secondary effects like the Sunyaev-Zel'dovich (SZ) effect 
\cite[]{aghanim99,cooray2001}, the Ostriker-Vishniac effect 
\cite[]{castro2002}, lensing \cite[]{cooray2001a,bernardeau2002} and others
add their own contributions to the total non-Gaussianity. To these
cosmological effects, we have
to add foregrounds as well as systematic effects and instrumental noise.
In order to disentangle all those sources from one another, it is essential
to assemble a ``tool-box'' of well-understood, fast and robust methods
for probing different aspects of the non-Gaussian signatures.

In this paper, we compare the behaviour of two different classes
of estimators, namely the higher order moments in wavelet space, and the 
bi- and trispectrum. Both of
these do not act directly on pixels, but are applied in a dual space.
The purpose of this study is not to present novel ways of detecting
non-Gaussian signatures, but to discuss practical aspects of these different
approaches, and to compare their behaviour when applied to a selection
of synthetic ``benchmark'' maps which are designed to mimic typical
non-Gaussian contributions to the CMB.
\par\bigskip
After the seminal work of \cite{pando98} which consisted of applying
wavelet techniques to look for non-Gaussian signatures in the COBE-DMR
data, several methods based on the wavelet
analysis have been developed in the context of the CMB detection of 
non-Gaussian signatures
\cite[]{aghanim99,forni99,hobson99,barreiro2001,cayon2001,jewell2001,martinez-gonzalez2002,starck2003}. These
methods have proven particularly suitable for statistical
studies due to the combination of two
characteristics. First, wavelets are quite localised both in space and
in frequency, which allows for the features of interest in an image to
be present at different scales. Second, the linear transformation
properties of Gaussian variables preserve Gaussianity. Conversely, any
non-Gaussian signal will exhibit a non-Gaussian distribution of its
wavelet coefficients. The two properties combined together allow us to
associate the statistical signatures with the spatial features that
have caused them. \par
The wavelet based estimators have been tested and qualified in terms
of their sensitivity to the non-Gaussian signatures on a variety of
simulated data sets such as cosmic strings, the SZ effect from galaxy
clusters, and anisotropies from inhomogeneous reionisation 
(see the above mentioned references). Also the
galactic contamination from dust emission
can induce foreground non-Gaussian signatures. Wavelets have been
used in such a context by \cite{jewell2001} to quantify the predicted
level of non-Gaussian features in the IRAS maps. Additionally, the
wavelet based non-Gaussian analysis has also been applied to the
COBE-DMR maps. \par
Different decomposition schemes and wavelet bases can be used for the
non-Gaussian studies.  In the following, we use 
a bi-orthogonal wavelet basis and a dyadic
decomposition scheme. This choice has been first motivated in
\cite{forni99} and \cite{aghanim99} by the fact that it is the best 
scheme in the
context of statistical analysis of CMB signals.  It is indeed optimal
for statistics since it gives, at each scale, the maximum number of
significant coefficients. In addition, it naturally allows us to
benefit from spatial correlations in the signal at each scale.  The
better performances of the bi-orthogonal dyadic wavelet decomposition
were confirmed by \cite{barreiro2001}, and also more recently by 
\cite{starck2003}, in a comparison with the ``\`a trous'' and ridgelet 
decompositions. \par\bigskip
We now focus on the Fourier based analysis.
The interest in using higher order correlation functions to study 
non-Gaussianity
dates back many years \cite[]{Luo94,jaffe94,heavens98}. The first detection
of a non-Gaussian contribution in real CMB data was achieved using the 
bispectrum \cite[]{ferreira98}, but later shown to be due to a weak 
systematic effect \cite[]{banday2000a}.

Numerous works deal with the construction of ideal 
higher order statistical estimators both in spherical and flat 
space as well as with their application 
to real data (CMB and large-scale structure survey data). 
In the CMB context, research focused on the two first higher order
correlation functions in Fourier space, namely the bispectrum 
(see recent work by~\cite{santos2002} and citations therein), 
and its 4-point equivalent, the trispectrum 
(see~\cite{hu2001a,kunz2001,cooray2002a}). 
An important advantage of $n$-point functions over other non-Gaussian
statistics such as wavelets or Minkowski functionals is that they 
are easier to predict for inflation but also for 
secondary sources. This makes them a very powerful test, 
explaining their interest for the scientific community.

Indeed, the Boltzmann equation, which describes the time evolution of
the perturbations in the cosmic fluid, does not mix different 
Fourier modes. The structure of linear perturbations is therefore generally
simple in Fourier space. Furthermore, the projection of the CMB
photons onto the sky sphere is simple as well, with the square
of the spherical Bessel functions coupling the Fourier mode $k$ and 
the angular scale, or more directly the multipole $\ell$, on the 
sphere. The success of studying the angular
power spectrum $C_\ell$ lies in that much of the initial conditions
is preserved throughout the evolution. Higher order correlation
functions can be treated in much the same way and allow for the use
of already existing techniques. It is therefore feasible to predict
theoretically the expected contributions to non-Gaussian signatures 
from different sources.
This allows us on the one hand to place limits on sources of
non-Gaussian features if none is detected, and on the other hand might
enable us to solve the inverse problem, namely identifying the origin
of the non-Gaussianity, if any is found.

In this work, we concentrate on the flat space approach. 
We use a normalised bispectrum probing all the possible triangle configurations
which exist in a homogeneous and isotropic universe,
and a normalised diagonal as well as nearly-diagonal trispectrum which 
depends on the side and the diagonal of the (nearly) trapezoidal configuration.
These are supposed to give us information on the scale dependence
of any non-Gaussian signal, but contrary to the wavelets, no 
spatial position information is retained.\par
We will start in the next section by presenting the data sets used for 
the comparison. In section 3, we examine the estimators of the non-Gaussian
signatures. The statistical analysis and the results are presented in section
4. We then end with our main conclusions.

\section{Data sets}
We test the statistical methods to detect non-Gaussian 
signals on three different data sets chosen so that they
are representative of many astrophysical situations. It is beyond 
the scope of the present study to characterise the astrophysical processes 
represented by the maps. The three sets have rather to be considered as test 
cases used in order to insure that the non-Gaussian detection is neither
specific to one peculiar type of non-Gaussian signatures, nor due to one 
particular detection method. 
Additionally, the combination of the different detection techniques on the
different sets of data allows us to compare the relative sensitivities of the
methods on an objective basis.\par
In practise, many of the astrophysical situations can be roughly classified 
into the following three different
types of signals: i) spherically shaped structures standing for compact 
astrophysical sources or beam-convolved point sources (e.g. galaxies, 
galaxy clusters) that we refer to as the point sources, ii) strongly 
anisotropic structures that represent filaments and elongated structures 
(e.g. interstellar clouds) that we refer to as the filaments, and iii)
non-linear couplings leading to non-Gaussian signatures, referred 
to as the $\chi^2$ (or $\chi$-squared) map, as we use 
the superposition of a Gaussian random field and its square.\par
In order to characterise the non-Gaussian signatures exhibited by all these
signals, the maps are compared to a set of 99 Gaussian realisations having 
the same power spectrum, referred to as the Gaussian reference set. Therefore, 
the only difference between Gaussian and non-Gaussian sets is due to the 
statistical nature of the two processes. All the maps (Gaussian and 
non-Gaussian) are composed of $500\times500$ pixels of 1.5 arc minutes aside.
\subsection{Filaments}\label{sec:iras}
\begin{figure}
\begin{center}
\includegraphics[width=\columnwidth]{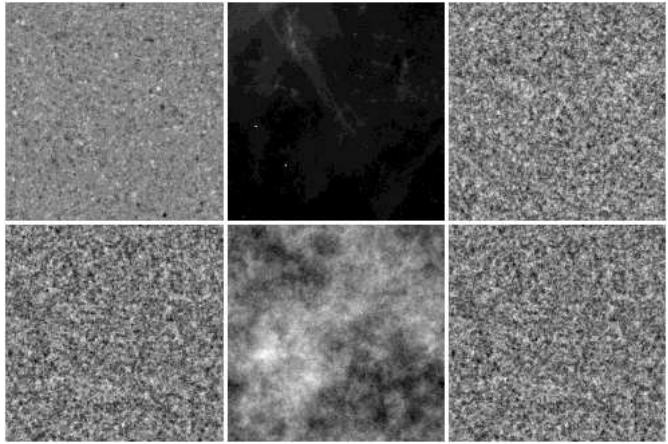}
\end{center}
\caption{Representative maps of the non-Gaussian signals and one of their
associated Gaussian realisations with the same power spectrum. Upper and lower 
left panels represent respectively a point source map and one Gaussian 
realisation. The
upper and lower middle panels are for the filaments and an associated Gaussian 
realisation. The upper and lower right panels represent the $\chi^2$ map and
a Gaussian field with the same power spectrum.}
\label{fig:maps}
\end{figure}

The set of maps for the filamentary structures is
a selection of 28 IRAS maps which represent the galactic dust emission at
high latitudes. These maps were chosen on the basis of their visual aspect 
so that they exhibit large and bright filaments (Fig. \ref{fig:maps}, upper 
middle panel). Therefore, they all have a
low amount of point sources and they exhibit elongated structures.
In order to focus on the
statistical properties of the maps, the selection has also been made on the
basis of their power spectrum. We choose the maps showing power spectra
with similar shapes regardless of their amplitudes. We then 
modify the initial IRAS maps so that their power spectra are rescaled.
Figure \ref{fig:cls}, lower panel, shows the power spectrum of the filaments.
Thanks to this rescaling, we can consider the 28 modified IRAS maps as 
statistical realisations of the same highly non-Gaussian process. This data 
set has a skewness (third moment) of 
$0.89 \pm 0.34$ ($1\,\sigma$), the excess kurtosis (fourth moment) is $2.63 \pm 
2.42$. The Gaussian reference maps (one example is given Fig. \ref{fig:maps}, 
lower middle panel) have a skewness of $-0.01 \pm 0.17$
and an excess kurtosis of $-0.11 \pm 0.19$.  
The large fluctuations of the skewness and excess kurtosis 
(also present in the Gaussian maps) arise due to the high power
on large scales associated with the large structures in the IRAS
maps. Hence, even the maps produced using a Gaussian random process could
be easily mistaken as being non-Gaussian, proving the necessity of comparing
a given test map with a set of Gaussian reference maps with the same power 
spectrum, rather than relying on a theoretical prediction of the Gaussian 
bounds. For a discussion of the Gaussian reference maps, see section 
\ref{sec:gaus}. 
\subsection{Point source maps} \label{sec:points}
\begin{figure}
\begin{center}
\includegraphics[width=\columnwidth]{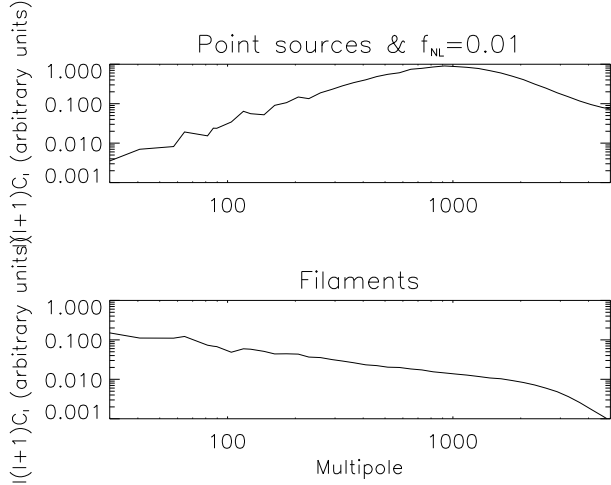}
\end{center}
\caption{The power spectra of the studied signals in arbitrary units. 
The upper panel shows the
power spectrum of both the point sources and the $\chi^2$ maps. The lower
panel represents the power spectrum of the filaments. }
\label{fig:cls}
\end{figure}

The second data set is referred to as the point sources. It is constituted 
of 50 simulated maps which consist of a distribution of Gaussian shaped 
sources having different sizes and amplitudes. They are randomly distributed
on the map and have positive and negative signs (see Fig. \ref{fig:maps}, left
upper panel). This signal is aimed at
reproducing the typical situation of a crowded field of 12.5 square degrees
containing $\simeq 10^7$ sources. The field therefore exhibits obvious 
confusion effects.
All the simulated maps have the same bell-shaped power spectrum (Fig. 
\ref{fig:cls}, upper panel) and exhibit a non-Gaussian signal. This 
non-Gaussian data set has a skewness of $-0.160 \pm 0.078$ and an excess kurtosis of 
$2.19 \pm 0.32$. A Gaussian reference set
of maps with the same power spectrum is simulated (Sect. \ref{sec:gaus}, Fig.
\ref{fig:maps}, left lower panel). The corresponding Gaussian maps have a 
skewness of $-0.003 \pm 0.022$ and an excess kurtosis of $-0.005\pm0.036$.
The point source maps are hence clearly marked as being non-Gaussian by
their excess of kurtosis.
\subsection{$\chi$-squared maps}\label{sec:chi}
The two previously described types of non-Gaussian signals 
are ideal to model astrophysical induced non-linearities due to foregrounds
and/or secondary effects. However in a wider context, we can expect 
the existence of primordial sources of non-Gaussian signatures.
A way of modelling some of these signatures is by introducing
a non-linear coupling in an originally Gaussian 
distributed perturbation field \cite[]{coles87,komatsu2000a,bartolo2002}.
The simplest weak non-linear coupling is given by 
$\chi({\mathbf x}) = \chi_L({\mathbf x}) 
                + f_{NL}(\chi_L^2({\mathbf x}) - <\chi_L^2({\mathbf x})>)$,
where $\chi_L$ denotes the original linear Gaussian field. 
The non-Gaussian signature introduced by the coupling is quantified 
by the so-called non-linear coupling constant $f_{NL}$ (note that 
$<\chi({\mathbf x})>=0$).

To generate these $\chi^2$ maps we use a method similar 
to the one followed to create a Gaussian
distributed field (see section \ref{sec:gaus}). The non-Gaussian character
is introduced by adding to the white noise 
field its squared minus its squared average weighted by $f_{NL}$. 
When multiplying the non-linear field in Fourier space with the
desired power spectrum, a normalisation by $1+2f^2_{NL}$ 
is needed in order to recover the power spectrum from the final map.   
In our approach, $f_{NL}$ quantifies the amount of the $\chi^2$ type
contribution in the Gaussian map.

We compute two sets of $50$ maps for 
$f_{NL}=0.01$ and $0.1$ in order to test the sensitivity
of the statistical methods to the amount of non-Gaussian signal 
introduced in the Gaussian field. For illustration purposes, we show in 
Fig. \ref{fig:maps} right upper and lower panels a
$\chi^2$ map for $f_{NL}=0.01$ and one associated Gaussian realisation.
The power spectrum of these maps was
chosen arbitrarily to be the same as for the point source maps.
As mentioned before, these maps are
not meant to reproduce accurately the physical processes -- to simulate 
a ``real'' CMB map with a
primordial $\chi^2$ contribution we would need to integrate over
the radiation transfer function. Nevertheless, to relate the order 
of magnitude
of our numbers to the ones found in the literature, we note that they
are rescaled by approximately a factor of $\Phi \sim \Delta T/T \sim
10^{-5}$ so that a value of $f_{NL} = 0.01$ here corresponds to about
$f_{NL} \approx 1000$ elsewhere (in reality a bit less, since e.g. on
large scales $\Delta T/ T \approx \Phi/3$). The $\chi$-squared maps 
with $f_{NL}=0.01$ have a skewness of $0.009 \pm 0.019$ and an excess
kurtosis of $0.002 \pm 0.031$, both of which are consistent with zero.
The maps with $f_{NL}=0.1$ on the other hand have a skewness
of $0.140 \pm 0.024$ and an excess kurtosis of $0.037\pm0.036$.
The Gaussian reference set is the same
as for the point source maps, see section \ref{sec:points}.
\subsection{Gaussian realisations} \label{sec:gaus}
In the process of investigating the statistical character of non-Gaussian 
signals, as we always deal with a finite ensemble of maps, there can
be substantial fluctuations in the results (see for example the
IRAS case, section \ref{sec:iras}). Furthermore, although not
important in our case but rather when analysing experimental data, 
often a ``Gaussian'' map is not created
directly from a pure Gaussian random process, but has super-imposed
known systematic effects (which
would result in spurious non-Gaussian features). In general,
the output of an estimator applied to a test-map is therefore not
compared to a theoretical result but to the corresponding output
obtained from a set of Gaussian reference maps (with
any known systematic effects one may need to add). Any detection of
non-Gaussian signatures can then be reliably quantified 
by comparing these two sets
of results, and we discuss this procedure in section \ref{sec:stat}
in more detail.

We generate two sets of Gaussian distributed fields with the 
power spectrum of the corresponding non-Gaussian set. The first
one is the aforementioned reference set. The second one (called
the Gaussian counterpart set) is strictly speaking redundant, but is used
to illustrate the fluctuations arising from finite ensembles, and
the presence of low-probability results which appear even when
comparing Gaussian maps with each other if many different tests
are applied.

The simplest standard method \cite[]{peacock99} 
is to create a spatial array of white noise using a random
number generator with zero mean and  unit variance. We can then 
give this Gaussian field the desired power spectrum by multiplying it
in Fourier space
with the square root of that power spectrum, the one
of the non-Gaussian set in our case. Using this procedure, we compute one 
set of 149 Gaussian maps (reference set and counterparts) with 
the same power spectrum as the 50 point-source maps, one set of 127 
Gaussian maps (reference set and counterparts) 
with the same power spectrum as the 28 IRAS modified maps. We arbitrarily 
choose the 
power spectrum of the point source maps for the $\chi^2$ maps and hence
use the same set of Gaussian realisations. We show
in Fig. \ref{fig:maps} (lower panels) one 
representative Gaussian realisation for each data set with the same spectrum 
as the non-Gaussian map. 
The skewness and the excess kurtosis of all the Gaussian reference sets and 
counterparts are expected to be zero. However, in practice they have a finite
scatter. The lower limit for the standard deviation of the skewness is given
for a normal distribution with unit variance by $\sqrt{15/N^2}$, or about
$0.0077$ in the case of $N^2=500^2$ independent values per map. The expected
standard deviation for the excess kurtosis, in the same idealised case, is 
$\sqrt{96/N^2}$ or $0.020$. One should note however that
the scatter of higher order moments is much larger in realistic cases
and is strongly susceptible to the details of the random process (see sections 
\ref{sec:iras}, \ref{sec:points} and \ref{sec:chi} for the actual numbers 
for the Gaussian reference sets). 
%
\section{Estimators of the non-Gaussian signatures}
We examine two large families of estimators for non-Gaussian signatures.
Both are based on dual space analysis. On the one hand, we focus on methods 
which are completely defined in Fourier space (namely the bi- and 
trispectrum). On the other hand, we study estimators in the wavelet space 
(namely third and fourth moments of the wavelet coefficients) 
in which signals are located both in pixel and in frequency space. It is
worth noting that the skewness (third moment) and excess kurtosis (fourth 
moment) of the wavelet coefficients are quite similar to 
respectively the bi- and trispectrum, in Fourier space. Additionally, 
we compute the cumulative probability function (CPF) of the pixel distribution.
This latter is also one of the Minkowski functionals, often called $\cal{A}$,
the surface of the excursion set at a given amplitude. As the CPF is a less
powerful non-Gaussian estimator, in the sense that it does not give 
information
on the scales nor on the physical location of any detection, we intend to use 
it only as a baseline for the detection of non-Gaussian signatures on 
a given map.

As discussed above, we analyse a certain number $N_{\mathrm NG}$ of 
non-Gaussian as well
as Gaussian maps with the same power spectrum and compare their properties
to a reference set of 99 Gaussian realisations, also with the same power 
spectrum. This section describes the statistical
estimators we use for our analysis. For a discussion of the tools used to 
interpret the results, see section \ref{sec:stat}.
\subsection{Pixel distribution function in real space}
As we have seen, in the Gaussian case each pixel is an independent
realisation of a Gaussian random process. Therefore, one of the easiest
and most direct tests of the Gaussian hypothesis just estimates the 
cumulative probability function (CPF) from the ensemble of pixels 
$N_{\rm tot}$. To this end, we set 
\be
P(<x) = N[{\rm pixels~with~values} < x] / N_{\rm tot}.
\ee
Since we have several independent maps for both the Gaussian as well
as the non-Gaussian processes, we can improve our estimate of the CPF,
and hence the sensitivity of the test, by just combining the pixels
from all maps. 
The two CPFs (from the Gaussian and non-Gaussian processes) can then be 
compared with statistical methods detailed in section \ref{sec:stat}.

Power spectra which vary extremely rapidly (like the ones used here)
can strongly amplify the random fluctuations. In
this case, the KS test will lead to spurious detections of non-Gaussianity 
even for Gaussian maps (see e.g. \cite{aghanim99}). It is therefore 
best to ``whiten'' the maps first, by deconvolving them with their 
estimated power spectrum. 

Due to its simplicity and its inherent locality (the CPF does
not test the relative position or configuration of the fluctuations,
as the location of the pixels in the map is not taken into account),
and as stated previously, we use the CPF method as a baseline for the 
comparison between the non-Gaussian methods. It should be stressed
clearly that this test can only detect non-Gaussian signatures but
does not give any further information about their location or scale 
dependence. It can therefore only very weakly distinguish between 
different non-Gaussian contributions.
\subsection{Wavelet space analysis}\label{sec:wdec}
The principle behind the wavelet transform
\cite[]{grossman84,daubechies88,mallat89} is to hierarchically decompose
an input signal into a series
of successively lower resolution reference signals and associated detail
signals. At each decomposition level, $j$, the reference signal has a
resolution reduced by a factor of $2^j$ with respect to the original signal.
Together with its respective detail signal, each scale contains the
information needed to reconstruct the reference signal at the next higher
resolution level.
\par
The wavelet analysis can be considered as a series of bandpass filters. 
It can
thus be viewed as the decomposition of the signal in a set of independent,
spatially oriented frequency channels. Using the
orthogonality properties, a function in this decomposition can be completely
characterised by the wavelet basis and the wavelet coefficients in this
decomposition. \par
The multi-level wavelet transform (analysis stage) decomposes the signal 
into sets of different frequency bands by iterative application of a 
pair of Quadrature Mirror Filters (QMF). A scaling
function and a wavelet function are associated with this analysis filter 
bank. The continuous scaling
function $\phi_{\mathrm A}(x)$ satisfies the following two-scale equation:
\begin{equation}
\phi_{\mathrm A}(x)= \sqrt{2} \sum_n h_{0}(n)\phi_{\mathrm A}(2x-n),
\label{eq1}
\end{equation}
where $h_{0}$ is the low-pass QMF.
The continuous wavelet $\psi_{\mathrm A}(x)$ is defined in terms of the
scaling function and the high-pass QMF $h_{1}$ through:
\begin{equation}
\psi_{\mathrm A}(x)=\sqrt{2} \sum_n h_{1}(n)\phi_{\mathrm A}(2x-n).
\end{equation}

The same relations apply for the inverse transform (synthesis stage) but,
generally, different scaling and wavelet functions ($\phi_{\mathrm S}(x)$ and
$\psi_{\mathrm S}(x)$) are associated with this stage:
\begin{equation}
\phi_{\mathrm S}(x)= \sqrt{2}\sum_n g_{0}(n)\phi_{\mathrm S}(2x-n),
\label{eq3}
\end{equation}
\begin{equation}
\psi_{\mathrm S}(x)= \sqrt{2}\sum_n g_{1}(n)\phi_{\mathrm S}(2x-n).
\end{equation}

Equations \ref{eq1} and \ref{eq3} converge to compactly supported basis
functions when
\begin{equation}
\sum_n h_{0}(n)=\sum_n g_{0}(n)=\sqrt{2}.
\end{equation}

The system is said to be bi-orthogonal if the following conditions are
satisfied:
\begin{eqnarray}
& & \int_{{R}}\phi_{\mathrm A}(x)\phi_{\mathrm S}(x-k)dx=\delta(k) \\
& & \int_{{R}}\phi_{\mathrm A}(x)\psi_{\mathrm S}(x-k)dx=0 \\
& & \int_{{R}}\phi_{\mathrm S}(x)\psi_{\mathrm A}(x-k)dx=0
\end{eqnarray}
\cite{cohen90} and \cite{vetterli95} give a
complete treatment of the relationship between the filter coefficients 
and the scaling functions.
\par
The wavelet functions are quite localised in space, and simultaneously they
are also quite localised in frequency. Therefore, this
approach is an elegant and powerful tool for image
analysis, because the features of interest in an image are present at
different characteristic scales. Consequently, different wavelet transforms
have been studied such as the `` \`a trous'' algorithm \cite[]{starck98}
that decomposes a $N \times N$ image $I$ as a superposition of
the form
\begin{equation}
I(x,y) = c_{J}(x,y) + \sum_{j=1}^{J} w_j(x,y),
\end{equation}
where $c_{J}$ is a coarse or smooth version of the original image $I$
and $w_j$ represents the details of $I$ at scale $2^{-j}$,
and the dyadic wavelet transform \cite[]{mallat98} that
decomposes the signal $s$ in a series of the form :
\begin{eqnarray}
 s(l) = \sum_{k} c_{J,k} (\phi_{\mathrm A})_{J,l}(k)
       +  \sum_{k} \sum_{j=1}^J (\psi_{\mathrm A})_{j,l}(k) w_{j,k}
\end{eqnarray}
where $J$ is the number of decomposition levels, $w_{j,k}$ the wavelet (or
detail) coefficients at position $k$ and scale $j$ (the indexing is such
that $j = 1$
corresponds to the finest scale, i.e. highest frequencies), and $c_{J}$ is
a coarse or smooth version of the original signal $s$. \par
In the present study and following \cite{forni99}, we choose
among all the possible bases and decomposition schemes the bi-orthogonal
wavelet basis and the dyadic decomposition. 
We have chosen to perform a six level ($J=6$) dyadic
decomposition of our data. In practice, a dyadic decomposition refers to 
a transform in which only the reference
sub-band is decomposed at each level. In this case, the analysis stage is
applied in both directions of the image at each decomposition level. The 
total number of sub-bands after $J$ levels of decomposition is then $3J+1$.
This decomposition gives as a result the details for the studied signal in
both directions. At each level, we therefore end up with 3 sub-bands that 
we refer to as the horizontal, vertical and diagonal details, plus the 
smoothed signal.  
This kind of decomposition furthermore allows us to benefit from correlations 
between the two directions at each level, and also from the maximum number
of coefficients, which is crucial for statistical analysis.
The linear transformation properties of Gaussian variables preserve the
statistical character making the wavelet coefficients
of a Gaussian process being Gaussian distributed. Conversely, any non-Gaussian 
signal will exhibit a non-Gaussian distribution of its wavelet coefficients. 
In addition, the multi-scale wavelet analysis allows us to associate the 
statistical signatures with the spatial features that have caused them.
However, we do not investigate the latter property in the present study.
\subsection{Fourier space analysis}\label{sec:fourier}
We now describe in some detail the Fourier space methods chosen
for this comparison. In Fourier space,
scalar functions on the sky sphere, such as the CMB temperature
fluctuations, are usually expressed as the coefficients
$a_{\ell m}$ of an expansion in spherical harmonics. For small patches,
we can instead approximate the sphere by a flat surface.
This makes it possible to use a fast Fourier transform to compute
the coefficients. In this case, the transformations between pixel
space (the original map $T(\bx)$) and Fourier space are given by
\bea
a(\bk) &=& \int \dx T(\bx) e^{2\pi i \bk \bx}
\rightarrow \sum_\bx T(\bx) e^{2\pi i \bk \bx /N^2} (\Delta x)^2 \\
T(\bx) &=& \int \dk a(\bk) e^{-2\pi i\bk \bx} \nonumber\\
&&\rightarrow \frac{1}{N^2}\sum_\bk a(\bk) e^{-2\pi i \bk \bx /N^2} (\Delta k)^2.
\eea
$N^2$ is the number of pixels per map. The flat space pixel size is 
determined from the angular pixel size $\theta$ (given in arc minutes) 
via $\Delta x = 2 \pi \theta / (360\times 60)$. The pixel size
in Fourier space is then $\Delta k = 1/\Delta x$ and the
angular scale in Fourier space is $\ell = 2\pi k$. 

The power spectrum of the temperature distribution is the two point
function, 
\be
C(k) = <a(\bk) a(-\bk)> \approx \int d\varphi a(k,\varphi)\, a(k,-\varphi) 
\label{defck}
\ee
The last correspondence is due to the statistical isotropy
of the temperature field on the sky sphere, which allows us
to replace the ensemble average with an average over directions
for a given mode.

For a Gaussian random field, all information is contained in the power
spectrum. All higher order $n$-point functions can be derived
from it. All functions with $n$ odd vanish, and the even $n$ ones
are found via Wick expansion. As an example, the four point
function is thus found to be $\langle a(k_1) a(k_2) a(k_3) a(k_4)\rangle 
= \langle a(k_1) a(k_2)\rangle \langle a(k_3) a(k_4)\rangle 
+\langle a(k_1) a(k_3)\rangle \langle a(k_2) a(k_4)\rangle 
+\langle a(k_1) a(k_4)\rangle \langle a(k_2) a(k_3)\rangle $, and so on.

The direct computation of higher order functions ($n>2$) for large maps
can be very slow. It generically scales as $N_{pix}^{n-1}$.
In the spherical case, a faster approach has been developed over the last
years \cite[]{spergel99,hu2001a,detroia2003}, by creating maps which 
contain only one scale $\ell$ each. We can use the same approach also in flat
space, defining the scale maps as
\be
T_\ell(\bx) = \frac{1}{N^2} \sum_\bk W_\ell(|\bk|) a(\bk) 
               e^{-2\pi i \bk \bx/N^2} (\Delta k)^2. 
\label{tldef}
\ee
The window functions $W_\ell$ serve to select the scale, generally
we will use $W_\ell(k) = 1$ if $k$ is in the band $\ell$ and $0$
otherwise. Other choices are clearly possible.

Although the formulation of the scale map method is more elegant
in spherical space, the flat space approach has the advantage of
using only sums, multiplications and fast Fourier transforms of
the map, rendering it very quick and efficient.

With this choice of scale maps, we can compute the power spectrum
via
\be
\sum_\bx T_\ell(\bx)^2 (\Delta x)^2 \sim
	\int dk\, d\varphi \, k\, W_\ell(k)\, a(k,\varphi)\, a(k,-\varphi)
\ee
which contains an additional weight $k$ compared to equation
(\ref{defck}). In the spherical case, we find instead a Wigner
3J symbol which has to be taken care of by the normalisation.
This corresponds in the flat space case to an infinitely narrow
bandwidth (i.e. window functions $W_\ell \propto \delta(2 \pi k-\ell)$). 
While a finite bandwidth 
will introduce a bias in the result, for sufficiently narrow bandwidths 
$k$ is approximately constant within the band and
just provides an offset which can also be taken care of via an
appropriate normalisation; in practice we just divide by the
number of pixels per band. Tests show that for a bandwidth
of two pixels we do not see any differences from eq. (\ref{defck})
even for rapidly varying power spectra. Similar normalisations
will have to be performed for the higher order correlation functions.

\subsubsection*{\bf The bispectrum}

The bispectrum is the three point function in Fourier space.
In the flat space approximation, it is given by a triangle
configuration,
\be
B_{\ell_1 \ell_2 \ell_3}
= <a(\bk_1) a(\bk_2) a(\bk_3)> \delta(\bk_1+\bk_2+\bk_3)
\ee
where the $\delta$ function arises due to statistical
isotropy (and all other configurations should vanish).

The above mentioned method of scale maps can be directly extended to 
higher moments. For the bispectrum, we compute
\bea
\lefteqn{\sum_\bx T_{\ell_1}(\bx) T_{\ell_2}(\bx) T_{\ell_3}(\bx) 
	(\Delta x)^2 =}\nonumber \\
&&\frac{1}{N^4}\sum_{\bk_1 \bk_2 \bk_3} a(\bk_1) a(\bk_2) a(\bk_3) \times \\
&& W_{\ell_1}(k_1) W_{\ell_2}(k_2) W_{\ell_3}(k_3) (\Delta k)^6\nonumber
\delta(k_1+k_2+k_3) .
\eea
Here, we normalised the $\delta$ symbol so that
$\delta(k=0) = 1/(\Delta k)^2$.

Again we use the approximation of narrow bands to derive
the normalisation. We assume that the bispectrum does not
vary strongly within a given band. Then the normalisation
which needs to be divided out is given by
\be
N(\ell_1,\ell_2,\ell_3) = \sum_{\bk,\bq} W_{\ell_1}(|\bk|)
W_{\ell_2}(|\bq|) W_{\ell_3}(|\bk+\bq|) ,
\ee
which we sum up using our choice of scale window functions.
Since this function is geometrical in nature and independent
of the actual map in question, it needs to be computed only
once. This is again just the number of configurations which
contribute to a given $(\ell_1,\ell_2,\ell_3)$ triplet.

The variance of this estimator for a Gaussian random field is
then found to be 
$\sigma_B^2 = 6 C_{\ell_1} C_{\ell_2} C_{\ell_3} /N(\ell_1,\ell_2,\ell_3)$ 
(for our $W_i$ which verify $W_i^2(\bk)
= W_i(\bk)$). 

We will in general divide out the dependence of the bispectrum
on the power spectrum, and we will do the same for the trispectrum.
This $C_\ell$ normalised estimator, $B_{\ell_1 \ell_2 \ell_3}/
\{C_{\ell_1} C_{\ell_2} C_{\ell_3}\}^{1/2}$ has comparable
properties to the unnormalised one, but is more robust with
respect to wrong estimations of the power spectrum. Komatsu et al. (2002)
even found it to have a slightly lower variance. 

\subsubsection*{\bf The trispectrum}

The trispectrum, the four point function, is slightly more complicated.
Statistical isotropy implies again a $\delta$-function over the four
momenta, so that we can visualise it
as a quadrilateral with sides $\ell_1$, $\ell_2$, $\ell_3$ and
$\ell_4$, and we can additionally specify (within limits) the length
of one of the diagonals, $a$. This additional degree of freedom
has to be taken into account when constructing an estimator.
To this end, we fix a diagonal $\ba$ with length $|\ba| = a$.
We then decompose the full trispectrum into two triangles,
averaged over all directions of $\ba$:
\be
T^a_{\ell_1,\ell_2,\ell_3,\ell_4}
= \sum_{\ba} \left(\tau_\ba^{\ell_1,\ell_2 *}\right) 
\tau_\ba^{\ell_3,\ell_4} W_a(|\ba|) (\Delta k)^2 + \mbox{permut.}
\ee

The sub-estimators $\tau_\ba^{\ell_i,\ell_j}$ need to be constructed 
so that one side
is given by $\ba$ and that the other sides have the lengths
indicated, e.g. $\ell_i$ and $\ell_j$. This defines the triangle
completely. A possibility for $\tau_\ba^{\ell_1,\ell_2}$ is
to use the Fourier transform of $T_{\ell_1}\cdot T_{\ell_2}$,
\be
\tau_\ba^{\ell_1,\ell_2} = 
\sum_\bx  T_{\ell_1}(\bx) T_{\ell_2}(\bx) e^{2\pi i \ba \bx / N^2} (\Delta x)^2 .
\ee
Then, the integral in the definition of the $T_\ell$ (see
eq.~(\ref{tldef})) over the exponentials leads to
$\delta(\bk_1+\bk_2-\ba) \delta(\bk_3+\bk_4+\ba)$ and
ensures the correct geometric structure.

In this study, we are not going to compute all possible trispectra. 
Instead, we are going to concentrate on two simpler cases.
The first one is the case $\ell_1 = \ell_2 = \ell_3 = \ell_4$,
\be
T^{(0)}_{\ell,a} \equiv T^a_{\ell,\ell,\ell,\ell}
\ee
This is quite a natural choice, and due to its symmetry well
suited to investigate elongated structures like filaments. Furthermore,
one needs only one scale at a given time and does therefore
not need to keep all scale maps in memory, greatly decreasing
the amount of memory needed by the algorithm.
But this ``diagonal'' estimator does not vanish in the case
of a Gaussian random field. It is in general proportional
to the square of the power spectrum. Specifically,
\be
\left<T^{(0)}_{\ell,a}\right>_G \propto
	C_\ell^2 \left( N(\ell)^2 \delta_{a,0} + 2 N(\ell,a)\right) .
\ee
Where the normalisation function for our choice of geometry
is given by
\be
N(\ell,a) = \sum_{\bk,\ba} W_a(|\ba|) W_\ell(|\bk|)
	W_\ell(|\ba+\bk|) ,
\ee
and again needs to be calculated only once. For $a = 0$
it is $N(\ell,a=0) = N(\ell) = \sum_\bk W_\ell(|\bk|)$,
the number of non-zero points in the scale map $\ell$
(for our choice of scale window functions $W$). This could
hide a real non-Gaussian contribution behind the Gaussian ``noise''.

It is also noteworthy that in this simple case the diagonal
is bounded by $0 \leq a \leq  2\ell$. 
As the length of
the other diagonal $b$ is related to a via $b=\sqrt{(4\ell^2-a^2)}$,
the cases with $a > \sqrt{2} \ell$ are redundant. But since the
binning in $a$ and $b$ is different, there is no easy way of
actually removing these cases. The canonical scale decomposition
in spherical space allows the orthonormalisation procedure described
below to remove the superfluous values in that case. 

We also use the near-diagonal case 
\be
T^{(+)}_{\ell,a} \equiv T^a_{\ell,\ell+1,\ell+2,\ell+3}
\ee
which vanishes (for $a\ne 0$) for a Gaussian random field.
 
As discussed in Kunz et al. (2001), the trispectrum $T^a_\ell$ is not unbiased.
This does not concern us overly much, since we only want to detect
non-Gaussianity in a first step, not estimate its precise amplitude.
Nonetheless, as discussed in that paper, the estimator given here
has many superfluous degrees of freedom which can influence the
degree of detection, depending on where precisely a signal is located
in $(\ell,a)$ space. We can decrease the number of estimators (and so
the superfluous degrees of freedom) as follows:
We define for each $\ell$ the matrix $M_{ab} = <T^a_\ell T^b_\ell>_G$, 
where $<>_G$
is the expectation value over a Gaussian ensemble, and diagonalise
it, $M_{ab} = \sum_i \lambda^i v_a^i v_b^i$. Since many of the
eigenvalues $\lambda^i$ vanish, we define a smaller ensemble
of estimators by the linear combinations
$\bar{T}^i_\ell = \sum_a (v_a^i/\sqrt{\lambda^{i}}) T^a_\ell$ if
$\lambda^{i} \neq 0$.

%
\section{Results}
\subsection{Statistical analysis} \label{sec:stat}
In the last section, we have described in great detail the estimators
which we apply to our synthetic data. Before we can discuss the
results, it is necessary to explain in which way we quantify
statistically the level of any non-Gaussian detection. To this
end, let us study each ``test'' separately. A ``test'' is {\em each} 
bi- or trispectrum configuration (as defined by the triplet or quadruplet
of $\ell$ values), and the skewness or excess kurtosis for the coefficients of 
each combination of scale and orientation in the wavelet case. Hence,
each of our four estimators is made up of many of these tests.
We apply each test independently to the non-Gaussian maps, their Gaussian 
counterparts 
as well as to the Gaussian reference set. We therefore end up with
three ensembles of values for each test.

\begin{figure}
\begin{center}
\epsfxsize=12cm
\epsfysize=10cm
\includegraphics[width=\columnwidth]{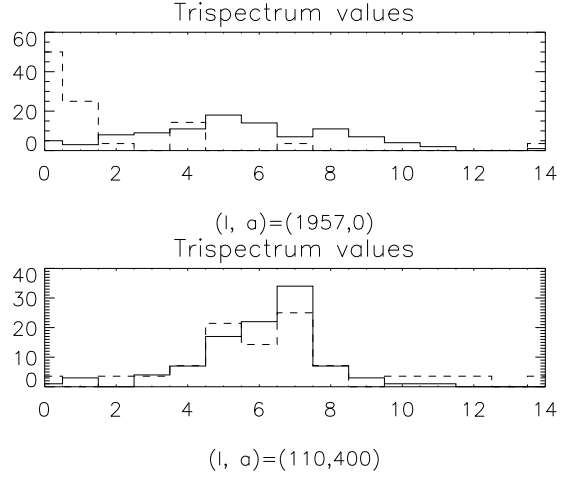}
\end{center}
\caption{Two distributions of near diagonal trispectrum values for 
the filaments: The
upper panel is for $(\ell,a)=(1957,0)$, a highly non-Gaussian case (dashed
line) and its corresponding values issued from the Gaussian set of maps 
(solid line). The lower panel is for $(\ell,a)=(110,400)$, a nearly Gaussian
case. In this case, the two distributions (same linestyle as upper panel) 
are very close to each other.}
\label{fig:dis-gng}
\end{figure}

We illustrate this by plotting in Fig. \ref{fig:dis-gng} (upper panel) two 
distributions of near-diagonal
trispectrum values $T^{(+)}_{\ell,a}$ with $(\ell,a)=(1957,0)$ for both the 
filament maps and the corresponding Gaussian reference maps. 
The lower panel shows the two distributions
for $(\ell,a)=(110,400)$. Clearly in the first case the two distributions
are very different, corresponding to a detection of non-Gaussianity.
It is this difference that needs to be quantified.
\par
A standard method to statistically compare two distributions 
is to perform a Kolmogorov-Smirnov (KS) test. 
The KS test returns the distance $d$ which
is defined as the maximum value of the absolute difference 
between the two cumulative distribution functions. One can derive the
probability $P_{KS}$ that the two data sets are drawn from the same
distribution (i.e. the two distributions are statistically the same)
using the following equation \cite[]{press92}:
\begin{equation}
P_{KS}=Q_{\mathrm KS}\left(\left[\sqrt{N_{\mathrm e}} +0.12+0.11/\sqrt{N_{\mathrm e}}\right]d\right),
\label{eq:ks}
\end{equation}
with 
\begin{equation}
Q_{\mathrm KS}(x)=2\sum_{j=1}^\infty(-1)^{j-1}\exp{(-2j^2x^2)}
\end{equation}
where $N_{\mathrm e}=\frac{N_1N_2}{N_1+N_2}$ is the effective number of data
points, given $N_1$ and $N_2$ the number elements of the two distributions.
In our study, we compare the distribution of values for a non-Gaussian
data set to a distribution obtained from the Gaussian reference set. 
Therefore, the values of the KS probabilities $P_{KS}$ represent a measurement 
of the confidence level, given by $1-P_{KS}$, for the detection of the 
non-Gaussian signatures. For example, if for a certain test $P_{KS}=10^{-3}$ 
then this test exhibits non-Gaussian features at a confidence level of 
$99.9\%$. The probabilities $P_G$ found by comparing the ensemble of 
Gaussian counterparts with the Gaussian reference ensemble are uniformly
distributed between $0$ and $1$. This, after all, is the definition
of the probability $P_{KS}$.

If we now move back from a single test to the estimators as a whole,
we are confronted with a new problem which is most noticeable in the
case of the bispectrum, for example. There, we perform of the order of 
$n_t\simeq 10^{5}$ tests. Hence, we expect to find some tests
(although very few) where even the comparison of two Gaussian ensembles 
results in a very low probability of order $1/n_t$. This is illustrated 
e.g. by figure \ref{fig:dist-bichi} where the uniform distribution of the
Gaussian probabilities is also clearly visible. Only a deviation from this 
behaviour
signals a violation of the Gaussian hypothesis. As a corollary, if a single 
test
(a single scale) has a probability $P_{KS} \ll 1/n_t$, that test alone ensures
with near-certainty that the process which created the map is not Gaussian.
To avoid any confusion with these small Gaussian probabilities, we state in
general both the results of the KS test of non-Gaussian versus Gaussian
reference maps as well as the Gaussian counterparts versus Gaussian reference
results.

Given the large ensemble of $P_{KS}$ values for the bi- and trispectrum,
and knowing their distribution in the Gaussian case (either taking the
theoretical uniform distribution or, which is more appropriate, directly 
the measured values),
it is tempting to go one step further and to apply the KS test a second
time. This {\em meta-statistics} then returns one 
{\em global meta-probability} for
the detection of a non-Gaussian signature by a given method, but neglects 
many aspects
such as correlations between the tests which make up that method. In the
case of the bi- and trispectrum, the {\em meta-statistics} 
discards the information on the frequency location of the tests and 
thus weakens the power of the Fourier analysis. In addition, the 
{\em meta-statistics} might also degrade 
our ability to detect the 
non-Gaussian signatures due to the known limitations of the KS
test. Even if, as an example, a few locations have very small
probabilities $P_{KS} \ll 1/n_t$, the {\em global meta-probability}
may be acceptable if those points are not numerous enough. 
Of course these highly significant values alone would suffice 
to reject the Gaussian hypothesis.

It is worth noting that the choice of the number of test (non-Gaussian) 
and reference maps for the first KS test has an important impact on the
{\em meta-statistics}. If the number of maps in the two sets have not enough 
different prime factors (e.g. 50 and 100), then heavy aliasing can occur 
since the KS test is only sensitive to the difference in the values of the 
cumulative distribution functions.
In this case, only very few distinct differences (and hence KS probabilities)
exist, leading to a non-continuous distribution of the KS probabilities.
As a consequence, the second KS test gives a spuriously low 
{\em global meta-probability} indicating that the signals are statistically
quite different, even when comparing Gaussian versus Gaussian maps. To avoid
this, we use 50 test maps and 99 reference maps, which largely eliminates
the problem.

For the wavelet based
estimators, we could also perform {\em meta-statistics}. We would in this case
have to create a distribution of KS probabilities for each decomposition
scale by comparing several sets (Gaussian versus Gaussian and non-Gaussian 
versus Gaussian), and then apply the KS test a second time.  
We refrain from applying this method here.

The KS test tends to be more sensitive around the median value of
the cumulative distribution function and less sensitive to differences at
the extreme ends of the distribution. It is therefore a good test to measure 
shifts but not so good in finding spreads which might affect the tails of 
the distribution. A way of accounting for these differences at the tails is 
to replace the KS distance $d$ by a stabilised or weighted statistics, 
for example the Andersen-Darling (AD) test or the Kuiper test. In the
former test, no probability is computed. This property makes it vital for the 
estimation of the non-Gaussian detection level that we actually 
compare on the one hand Gaussian counterparts and the Gaussian reference
set, to on the other hand non-Gaussian data set and the reference set.
We have checked that both the AD and the Kuiper tests give results that are
in quite agreement with the KS test. In the following, we concentrate onto 
the KS test, and we present the results for each statistical estimator in
terms of the probability that two distributions are identical.
\subsection{Results from the cumulative probability function} \label{sec:res-cpf}
As mentioned before, our baseline is given by a simple comparison
between the estimated CPF of the non-Gaussian data sets and the
corresponding reference set. We use the KS test for this comparison,
since it provides us directly with an estimate of the probability.
We do not collect all pixels of all
test maps, but use only the pixels of one map at a time and find
therefore a distribution of values.
Table \ref{tab:ks-cpf} gives the averaged results obtained in this way, 
compared to
the results from the  Gaussian counterpart set. The maps used for
this purpose were deconvolved with the average power spectrum
extracted from the comparison set.

In all cases, the non-Gaussian signature is clearly
detected (with the $f_{NL} = 0.01$ $\chi^2$ type maps being the least
non-Gaussian by far), while the Gaussian counterpart maps are consistent with
the Gaussian hypothesis. As the point source maps and the $\chi^2$ maps use
the same Gaussian reference and counterpart maps, they also share
the same Gaussian probability $P_G$. That value might be slightly
higher than expected due to the deconvolution which has to use an
estimated power spectrum (the true one being unknown in practice). 
But in the worst case, we expect this measure to lead to a weaker detection
of non-Gaussianity. If we do not deconvolve the
maps, we find, $P_{G} = 10^{-3}$ for the point source Gaussian reference 
set versus the Gaussian counterpart set, and the filament case is much worse.

\begin{table}
\begin{center}
\begin{tabular}{|c|ccc|}
\hline
& Filaments   & Point sources  & $\chi^2$ maps   \\
\hline
$P_{KS}$ & $0$     & $0$     & $4.7\cdot10^{-4}$      \\
$P_{G}$  & $0.73$ & $0.72$ & $0.72$  \\
\hline
\end{tabular}
\end{center}
\caption{Test for non-Gaussianity using the CPF for all (whitened) maps. 
The first line is the average KS 
probability for the real, non-Gaussian maps,
the second line the average probability for the Gaussian counterpart set. The 
$\chi^2$ type maps have a non-linear coupling factor $f_{NL}=0.01$.}
\label{tab:ks-cpf}
\end{table}

Additionally, we can quantify the sensitivity of this test to the
non-Gaussian signatures present in the $\chi^2$ maps. To this end, we form
a likelihood estimator ${\cal L}[f_{NL}]$, and find the best-fit
parameter $f_{NL}$ by minimising ${\cal L}$. The variation of the
recovered $f_{NL}$ for a set of Gaussian maps gives an indication 
of the sensitivity of the approach.
A priori, we could use directly the KS probability as given above
for the likelihood function. But, as demonstrated in \cite{santos2002},
it is simpler to use a $\chi^2$ fitting procedure to a signal which
can be analytically predicted and in which $f_{NL}$ enters linearly. 
In this case, we can calculate the best-fit value in a very straightforward
way.

\begin{figure}
\includegraphics[width=\columnwidth]{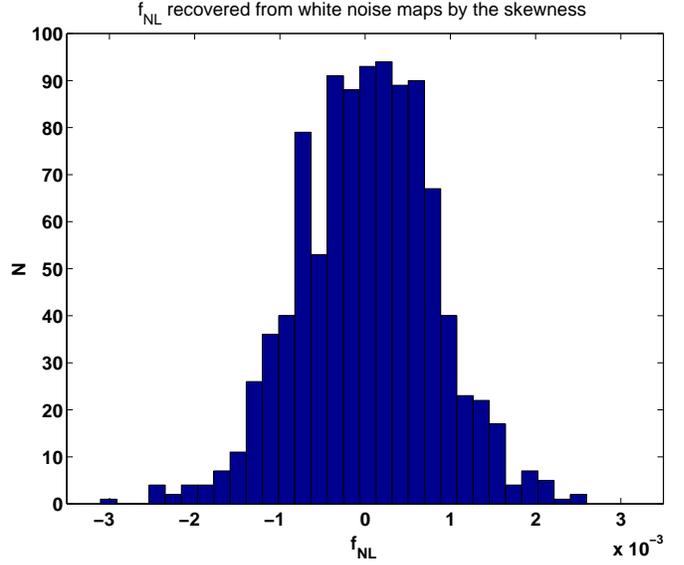}
\caption{The values of $f_{NL}$ recovered from the skewness of 1000 maps
which contain Gaussian white noise. The variance is $\Delta f_{NL}=8.\, 10^{-4}$. }
\label{fig:cpf_fnl}
\end{figure}

The probably simplest signals which can be extracted from the CPF
are its moments. As the average vanishes, and as all even moments
show only a leading order non-Gaussianity proportional to $f_{NL}^2 \ll 1$,
it is best to use the skewness. To first order in
$f_{NL}$, we find $<T^3> = 6 f_{NL} \sigma^4$, where the standard
deviation $\sigma$ is the one of the original Gaussian map. The
one recovered from a non-Gaussian $\chi^2$ type map differs by a term
of order $f_{NL}^2$ which can be neglected. The likelihood is then
given by
\begin{equation}
\chi^2[f_{NL}] = \sum_\bx 
\frac{\left(T(\bx)^3 - f_{NL} 6 \sigma^4\right)^2}{15 \sigma^6} .
\label{eq:cpf-fnl}
\end{equation}
Since it is again important
to use a whitened map we will assume that $\sigma = 1$.
We set $\partial \chi^2 / \partial f_{NL} = 0$ and find
\begin{equation}
f_{NL} = \frac{1}{6 N^2} \sum_\bx T(\bx)^3 .
\end{equation}
Figure \ref{fig:cpf_fnl} shows the recovered $f_{NL}$ from 1000
maps with Gaussian white noise. Their variance is $\sigma = 8.\, 10^{-4}$
and we find for the 50 maps with $f_{NL} = 0.01$ that
$f_{NL} = 0.0098 \pm 0.0007$. 

There is a caveat: In order to extract
that value, we needed to know the exact kind of non-Gaussian signal we are
dealing with. However, 
it is valid to use the present procedure to set limits on the parameter in
question if {\em no} non-Gaussian signal is detected.

\subsection{Results from the wavelet decomposition}
In order to find the non-Gaussian signatures present in the studied 
signals, we compute the skewness and excess kurtosis
of the wavelet coefficients at each decomposition scale and 
in each sub-band, i.e for the three types of
details (horizontal, vertical and diagonal). We compute these quantities 
for both the Gaussian and non-Gaussian maps and we compare the obtained 
distributions of skewness and excess kurtosis
for the Gaussian and non-Gaussian processes using the KS test. 
This comparison allows us to get, for each scale and orientation, the
probability that the non-Gaussian process and the Gaussian realisations 
with the same power spectra have the same distribution. \par
All the obtained probabilities for the skewness and excess kurtosis
are given in the tables \ref{tab:ks-khi001}, \ref{tab:ks-iras} and
\ref{tab:ks-poi}. In these tables and for each type of details, the first
set of 2 lines represents $P_{KS}$ for the skewness. The first line is for
non-Gaussian versus reference set comparison, and the second line is for 
the Gaussian
counterpart versus reference set (G vs G). The second set of 2 lines 
represents the $P_{KS}$ for the excess kurtosis. Again, the first line is for
non-Gaussian versus reference set comparison, and the second for the Gaussian
counterpart versus reference set.\\
The probabilities of the order of, or lower than,
$10^{-3}$ indicating a non-Gaussian detection at a level $>99.9\%$ are
all in shaded boxes. The probabilities obtained from the
comparison between the Gaussian counterparts and the reference Gaussian
realisations (G vs G) allow us to estimate any possible non-significant 
detection of non-Gaussian signature, or statistical fluctuation in the 
Gaussian sets.
\subsubsection{Third moment of the wavelet coefficients: Skewness}
Confirming the CPF results for which there was a clear detection
of non-Gaussian signatures in the set of $\chi^2$ maps 
with a coupling factor $f_{\mathrm{NL}}=0.1$, the non-Gaussian 
signatures are very easily detected in the wavelet space. We therefore focus on
the $\chi^2$ maps with a coupling factor $f_{\mathrm{NL}}=0.01$.
We summarise the KS probabilities derived from the comparison of the
distribution of skewnesses in Table \ref{tab:ks-khi001}, first two lines of
each type of details. 
\begin{table*}
\begin{center}
\begin{tabular}{|c|cccccc|}
\hline
& Scale 1 & Scale 2  & Scale 3 & Scale 4 & Scale 5 & Scale 6  \\
\hline
Vertical &  \colorbox{light}{2.08$\,10^{-18}$}& 0.226& 0.488& 0.828& 0.158& 0.368 \\
G vs G  & 0.411& 0.450& 0.160& 0.160& 0.791& 0.036  \\
\hline
        & 0.488& 0.426& 0.367& 0.426& 0.426& 0.189 \\
G vs G  &0.332 & 0.791& 0.068& 0.940& 0.411& 0.332 \\
\hline
\hline
Horizontal& \colorbox{light}{1.97$\,10^{-17}$}& 0.002& 0.828& 0.555& 0.695& 0.625  \\
G vs G  & 0.876& 0.207& 0.940& 0.876& 0.940& 0.332\\
\hline
        & 0.315& 0.0450& 0.625& 0.764& 0.426& 0.226\\
G vs G  & 0.264& 0.596& 0.411& 0.940& 0.596& 0.791\\
\hline
\hline
Diagonal & \colorbox{light}{0.001}& 0.157& 0.828& 0.368& 0.315& 0.315 \\
G vs G  & 0.791& 0.596& 0.940& 0.695& 0.049& 0.791 \\
\hline
         & 0.828& 0.695& 0.930& 0.828& 0.764& 0.963\\
G vs G   & 0.265& 0.265& 0.876& 0.791& 0.499& 0.596\\
\hline
\hline
\end{tabular}
\end{center}
\caption{For the $\chi^2$ maps with $f_{\mathrm NL}=0.01$: The KS probability 
(for two signals to have identical distributions) for
each of the details at all decomposition scales. For each detail, the first 
two lines are for the skewness of the wavelet coefficients, and the second
two lines represent the excess kurtosis. For each pair, the first line stands for 
the comparison between the 50 non-Gaussian and 99 reference Gaussian maps, 
and the second line for the comparison between the 50 Gaussian counterparts 
and 99 reference Gaussian maps (G vs G). All probabilities lower than 
$10^{-3}$ are in shaded boxes.}
\label{tab:ks-khi001}
\end{table*}
For these $\chi^2$ maps, the departure from the Gaussian hypothesis 
is only detected at the first wavelet decomposition scale
(i.e 3 arc minutes). Both the vertical and horizontal details appear very
sensitive to the non-Gaussian features present in the maps. The non-Gaussian
signatures are also detected in the diagonal details, at the first scale, 
but with a much lower confidence level. From the KS probabilities
obtained from the comparison Gaussian counterparts versus Gaussian reference 
maps (second line of the first set in each detail), it is obvious that
the detection of the non-Gaussian signatures are highly significant. It is
however worth noting that we detect some fortuitous non-Gaussian signatures,
at 93 to 96\% confidence levels, due to statistical fluctuations
when we compare the Gaussian maps with the reference Gaussian set.
 \par\bigskip
In the case of the filaments (Table \ref{tab:ks-iras}, first
two lines of each type of detail), the KS test for the skewness distribution 
of the non-Gaussian against the Gaussian reference set
shows that the non-Gaussian features are detected at all scales (with less
confidence at scale 6). The detection takes place for the three types of
details. We note that the probability for the non-Gaussian signal to be
compatible with a Gaussian process increases for the highest decomposition 
levels, i.e. the largest angular scales, as it is expected from our studied 
signal in which the filamentary structures have not increasingly large 
sizes.\par\bigskip
As for the point sources (Table \ref{tab:ks-poi}, first two lines of
each type of details), the non-Gaussian signatures 
do not show up very significantly (whatever decomposition scale and details).
The obtained probabilities remain always larger than $\simeq 0.002$
suggesting a non-Gaussian detection at a $\simeq 99.8\%$ confidence level 
at best. 
\par
For the filamentary structures as well as for the point sources, the 
comparison between Gaussian counterparts and Gaussian reference (G vs G) 
set shows that we are very much compatible 
with identity test (processes having the same distribution). The obtained 
KS probabilities show the high level of significance of the non-Gaussian 
detection.
\begin{table*}
\begin{center}
\begin{tabular}{|c|cccccc|}
\hline
& Scale 1 & Scale 2  & Scale 3 & Scale 4 & Scale 5 & Scale 6  \\
\hline
Vertical& \colorbox{light}{1.27$\,10^{-7}$}& \colorbox{light}{6.54$\,10^{-15}$}& \colorbox{light}{1.77$\,10^{-11}$}& \colorbox{light}{1.65$\,10^{-8}$}& \colorbox{light}{5.25$\,10^{-8}$}& 0.047\\
G vs G   & 0.743& 0.911& 0.486& 0.879& 0.320& 0.239 \\
\hline
 & \colorbox{light}{3.3$\,10^{-20}$}& \colorbox{light}{3.3$\,10^{-20}$}& \colorbox{light}{3.3$\,10^{-20}$}& \colorbox{light}{3.3$\,10^{-20}$}& \colorbox{light}{1.01$\,10^{-16}$}& \colorbox{light}{1.09$\,10^{-6}$}\\
G vs G        & 0.379& 0.574& 0.471& 0.996& 0.682& 0.879\\
\hline
\hline
Horizontal& \colorbox{light}{1.77$\,10^{-11}$}& \colorbox{light}{1.43$\,10^{-12}$}& \colorbox{light}{1.9$\,10^{-9}$}& \colorbox{light}{1.77$\,10^{-11}$}& \colorbox{light}{9.23$\,10^{-8}$}& \colorbox{light}{0.001}\\
G vs G        & 0.651& 0.788& 0.976& 0.391& 0.515& 0.998 \\
\hline
& \colorbox{light}{3.3$\,10^{-20}$}& \colorbox{light}{3.3$\,10^{-20}$}& \colorbox{light}{3.3$\,10^{-20}$}& \colorbox{light}{3.3$\,10^{-20}$}& \colorbox{light}{1.22$\,10^{-18}$}& \colorbox{light}{7.68$\,10^{-9}$}\\
G vs G        & 0.773& 0.066& 0.829& 0.258& 0.088& 0.559 \\
\hline
\hline
Diagonal& \colorbox{light}{5.13$\,10^{-7}$}& \colorbox{light}{2.85$\,10^{-5}$}& \colorbox{light}{1.65$\,10^{-8}$}& \colorbox{light}{0.0004}& 0.010& 0.008\\
G vs G        & 0.682& 0.142& 0.040& 0.651& 0.842& 0.682 \\
\hline
& \colorbox{light}{3.3$\,10^{-20}$}& \colorbox{light}{3.3$\,10^{-20}$}& \colorbox{light}{3.3$\,10^{-20}$}& \colorbox{light}{3.3$\,10^{-20}$}& \colorbox{light}{3.7$\,10^{-16}$}& \colorbox{light}{5.69$\,10^{-8}$}\\
G vs G        & 0.636& 0.529& 0.758& 0.816& 0.379& 0.040\\
\hline
\hline
\end{tabular}
\end{center}
\caption{For the filaments: The KS probability (for two signals to have
identical distributions) for
each of the details at all decomposition scales. For each detail, the first 
two lines are for the skewness of the wavelet coefficients, and the second
two lines represent the excess kurtosis. For each pair, the first line stands for 
the comparison between the 28 non-Gaussian and 99 reference Gaussian maps, 
and the second line for the comparison between the 28 Gaussian counterparts 
and 99 reference Gaussian maps (G vs G). All probabilities lower than 
$10^{-3}$ are in shaded boxes.}
\label{tab:ks-iras}
\end{table*}
\subsubsection{Fourth moment of the wavelet coefficients: Excess kurtosis}
We now turn to the fourth moment in the wavelet space, i.e. the excess kurtosis
of the wavelet coefficients. The results are given by the second sets of two
lines in Tables \ref{tab:ks-khi001}, \ref{tab:ks-iras} and \ref{tab:ks-poi}.
Again the first line is for the KS probability obtained from the comparison
of non-Gaussian maps with Gaussian reference set, and the second line stands
for the comparison of Gaussian counterparts versus reference set (G vs G).

In the case of the $\chi^2$ maps with $f_{\mathrm NL}=0.01$,
the fourth moment do not, as expected, exhibit any departure from the 
Gaussian hypothesis.
This is demonstrated by the large values of the KS probabilities (Table 
\ref{tab:ks-khi001}, second sets of lines).\par\bigskip
The point sources (Table \ref{tab:ks-poi}, first line of the second set of 
lines for each detail), exhibit very highly non-Gaussian
signatures as shown by the small KS probabilities 
(always smaller than $\simeq 10^{-5}$). Furthermore, the non-Gaussian
signatures are present at all decomposition scales and for all the details.
We compare the obtained probabilities with those derived from the
comparison of Gaussian counterparts with Gaussian reference set. We note
the high level of significance of the non-Gaussian detections.
\par\bigskip
The same kind of results (detection of non-Gaussian features at all
scales with all details) are obtained for the filaments
(Table \ref{tab:ks-iras}, second sets of lines) for which the
probability that the non-Gaussian maps are statistically equivalent to
the Gaussian reference maps is always smaller than $10^{-6}$. We have
again computed the KS probabilities for the excess kurtosis of the wavelet
coefficients from the comparison of Gaussian maps. This comparison
shows that the two sets of maps are statistically the same which
conversely reinforces the level of significance of the non-Gaussian
detection in the filament maps.
\begin{table*}
\begin{center}
\begin{tabular}{|c|cccccc|}
\hline
& Scale 1 & Scale 2  & Scale 3 & Scale 4 & Scale 5 & Scale 6  \\
\hline
Vertical& 0.012& 0.791& 0.002& 0.008& 0.411& 0.791\\
G vs G        & 0.791& 0.265& 0.122& 0.332& 0.791& 0.160 \\
\hline
 & \colorbox{light}{3.47$\,10^{-30}$}& \colorbox{light}{3.47$\,10^{-30}$}& \colorbox{light}{3.47$\,10^{-30}$}& \colorbox{light}{6.12$\,10^{-24}$}& \colorbox{light}{0.001}& 0.411\\
G vs G        & 0.265& 0.876& 0.207& 0.999& 0.499& 0.160 \\
\hline
\hline
Horizontal& 0.122& 0.695& 0.265& 0.006& 0.265& 0.791\\
G vs G        & 0.596& 0.411& 0.876& 0.411& 0.499& 0.596 \\
\hline
& \colorbox{light}{3.47$\,10^{-30}$}& \colorbox{light}{3.47$\,10^{-30}$}& \colorbox{light}{3.47$\,10^{-30}$}& \colorbox{light}{2.3$\,10^{-22}$}& \colorbox{light}{1.42$\,10^{-6}$}& 0.068\\
G vs G        & 0.411& 0.499& 0.122& 0.791& 0.695& 0.596 \\
\hline
\hline
Diagonal& 0.012& 0.160& 0.596& 0.596& 0.876& 0.332\\
G vs G        & 0.695& 0.996& 0.940& 0.876& 0.207& 0.160 \\
\hline
& \colorbox{light}{3.47$\,10^{-30}$}& \colorbox{light}{3.47$\,10^{-30}$}& \colorbox{light}{1.09$\,10^{-26}$}& \colorbox{light}{9.87$\,10^{-16}$}& \colorbox{light}{1.02$\,10^{-7}$}& \colorbox{light}{0.0004}\\
G vs G        & 0.092& 0.160& 0.876& 0.791& 0.596& 0.499 \\
\hline
\hline
\end{tabular}
\end{center}
\caption{For the point sources: The KS probability (for two signals to have
identical distributions) for
each of the details at all decomposition scales. For each detail, the first 
two lines are for the skewness of the wavelet coefficients, and the second
two lines represent the excess kurtosis. For each pair, the first line stands for 
the comparison between the 50 non-Gaussian and 99 reference Gaussian maps, 
and the second line for the comparison between the 50 Gaussian counterparts 
and 99 reference Gaussian maps (G vs G). All probabilities lower than 
$10^{-3}$ are in shaded boxes.} 
\label{tab:ks-poi}
\end{table*}
\subsection{Results from the bispectrum}
Similarly to the wavelet coefficients, we compare for each triplet
$(\ell_1,\ell_2,\ell_3)$ the distributions of the (normalised) bispectrum 
values, as described in section~\ref{sec:fourier}, for the
non-Gaussian realisations and for the Gaussian reference set. Hence, we
associate to each triplet
$(\ell_1,\ell_2,\ell_3)$ two ensembles of bispectrum coefficients --
on the one hand 50 or 28 values for the non-Gaussian maps, and on the 
the other hand 99 values for the Gaussian reference set. 
Using the KS test, we therefore
end up for each triplet with a probability that the signals are drawn from
the same process, i.e. that the non-Gaussian maps are compatible with 
the Gaussian reference set. 

For the full bispectrum, we limit ourselves to a bandwidth of four pixels,
but probe additionally the diagonal elements with a bandwidth of two pixels.
We have tested that using the higher resolution for the full bispectrum
does not lead to a better detection
of the non-Gaussian signal for the $\chi^2$ type maps, and does not improve
the variance of the recovered value of $f_{NL}$ (see the following subsection).

For the sake of a clearer visual representation
of the results, we choose to show the KS probabilities as a function of the
surface and the smallest angle defined by the triplet, rather than as a
function of the triplet itself. Note that we are speaking of surfaces 
and angles in Fourier space. 
\subsubsection{$\chi^2$ maps}
\begin{figure}
\includegraphics[width=\columnwidth]{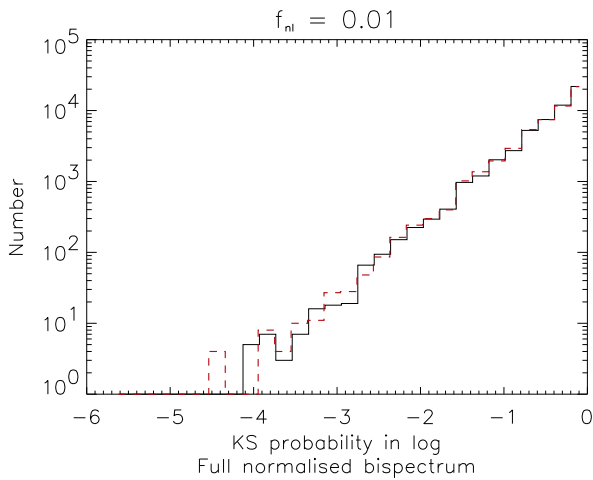}
\caption{For the $\chi^2$ maps with non-linear coupling factor $f_{NL}=0.01$: 
Distribution of the KS probabilities obtained by i) comparing the 
full normalised bispectrum estimator of the non-Gaussian 
maps to the Gaussian reference set (dashed line), and ii) comparing the same
quantity for the Gaussian counterparts to the Gaussian reference set 
(solid line). In the second case, the probabilities are distributed uniformly. The
slope is due to the log-log representation we have chosen.}
\label{fig:dist-bichi}
\end{figure}

We first focus on the case of the non-Gaussian $\chi^2$ maps with a
coupling factor of $f_{\mathrm{NL}}=0.01$. 
We find, for the full bispectrum, that the probabilities for the
non-Gaussian signal to be compatible with its Gaussian reference set are 
as low as a few times $10^{-6}$. However as mentioned in 
section \ref{sec:stat} (see also Fig. \ref{fig:dist-bichi}), 
the comparison of the 50 Gaussian counterpart realisations to the 
reference set of 99 Gaussian maps gives KS probabilities of the same order,
and it is hard to say whether there is a detection of non-Gaussianity
or not. If we apply the KS test
a second time, now to the two resulting distributions of probabilities,
the {\em global meta-probability} obtained is $4.1\,10^{-4}$. The bispectrum
therefore does indeed detect this type of non-Gaussian signatures.

\begin{figure}
\includegraphics[width=\columnwidth]{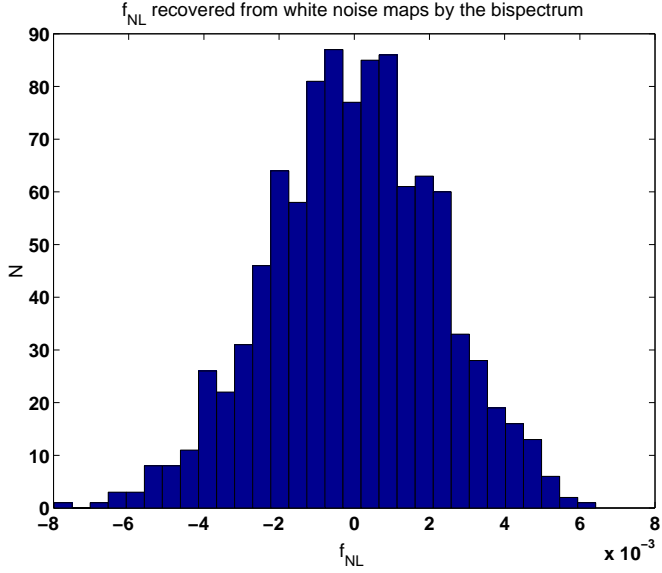}
\caption{The values of $f_{NL}$ recovered from the bispectrum of 1000 maps
which contain Gaussian white noise. The variance is 
$\Delta f_{NL} =2.\,10^{-3}$. }
\label{fig:bi_fnl}
\end{figure}

As the bispectrum shares many properties with the skewness, we would expect it
to be rather good at detecting $\chi^2$ type non-Gaussianity. Given the
way in which we construct our test maps, the theoretical signal is
\begin{equation}
B^{(th)}_{\ell_1 \ell_2 \ell_3} = 6 f_{NL} \sqrt{ C_{\ell_1} C_{\ell_2} C_{\ell_3}} .
\label{eq:bi-th}
\end{equation}
We can eliminate any dependence on the power spectrum by using the $C_\ell$-
normalised estimator.
The $\chi^2$ likelihood function is then (denoting a triplet 
$(\ell_1, \ell_2, \ell_3)$ by $\alpha$) very similar to the one
for the CPF, eq.~(\ref{eq:cpf-fnl}), 
\begin{equation}
\chi^2[f_{NL}] = \sum_\alpha  \frac{\left( B_\alpha^{(obs)} - 6 f_{NL} \right)^2}
{\sigma_\alpha^2}
\end{equation}
and since the dependence of $\partial \chi^2 / \partial f_{NL} = 0$ on
$f_{NL}$ is linear we can again solve for it analytically.
By using this method, we recover $f_{NL} = 0.010 \pm 0.002$ and find that
the 1000 Gaussian white noise maps have a variance of $0.0023$ (see 
Fig.~\ref{fig:bi_fnl}).
As mentioned earlier, using a bandwidth of two instead of four does not
improve the sensitivity. One should not forget, though, that the Fourier
space methods are better adapted to spherical spaces, where the scale 
decomposition
is automatic and does not need to be imposed by placing rings on a rectangular
grid. Furthermore, the skewness is always just a number, while a realistic 
model of primary non-Gaussian signatures (taking into account the radiation 
transfer function) introduces a non-trivial shape and even sign changes into 
eq.~(\ref{eq:bi-th}), see e.g.~\cite{santos2002}. In the presence of several
different kinds of non-Gaussian signatures, the bispectrum should be better 
at discriminating between them in that case, and also using a lower bandwidth 
may become useful.
\subsubsection{Point sources}
\begin{figure}
\begin{center}
\includegraphics[width=\columnwidth]{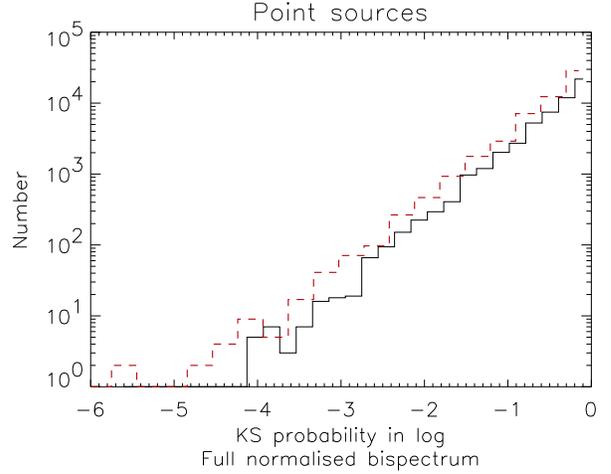}
\end{center}
\caption{For the point sources: Distribution of the KS probabilities for 
the full bispectrum. The dashed line represents the results from the comparison
of non-Gaussian maps to Gaussian reference set. The solid line is for the 
comparison Gaussian counterparts versus reference set.}
\label{fig:dist-bipt}
\end{figure}

In the case of the non-Gaussian maps made of a distribution of point sources,
the KS test gives again probabilities down to $10^{-6}$ for the full 
bispectrum, apart from a single point with much lower probability 
($P_{KS}\simeq 10^{-9}$).
We concentrate onto the full bispectrum tests and we need to 
compare the obtained KS probabilities with the ones derived from the 
comparison of 50 Gaussian counterparts versus the reference set.
The figure \ref{fig:dist-bipt} displays the distribution 
of KS probabilities. We note in particular that the non-Gaussian signal 
exhibits an excess
of low KS probabilities as compared with the probabilities from Gaussian vs 
reference set. This difference between the two distributions 
is confirmed by the {\em meta-statistics} which gives a probability 
of $8.6\,10^{-4}$.\par
\begin{figure}
\begin{center}
\includegraphics[width=\columnwidth,height=7cm]{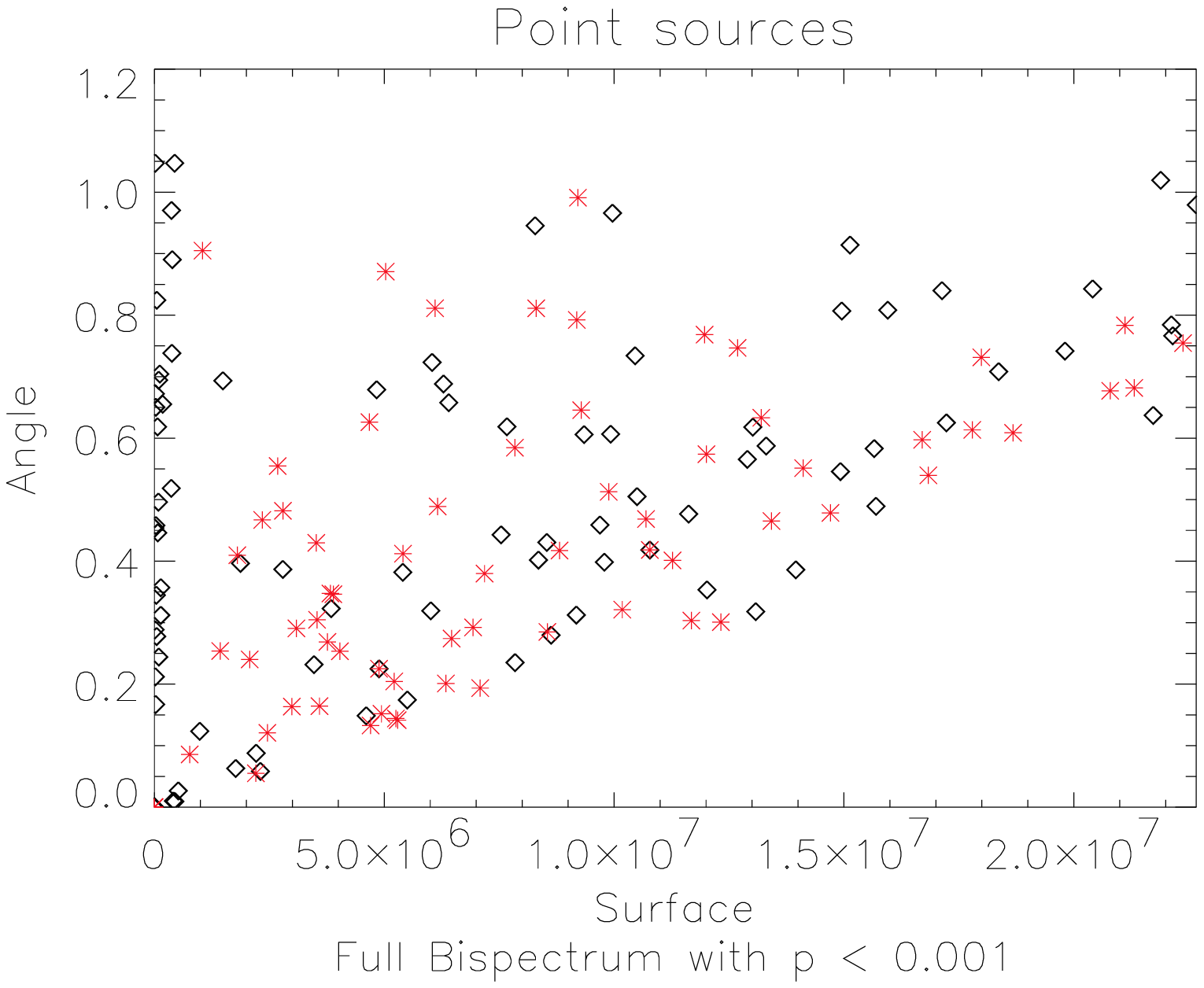}
\includegraphics[width=\columnwidth,height=7cm]{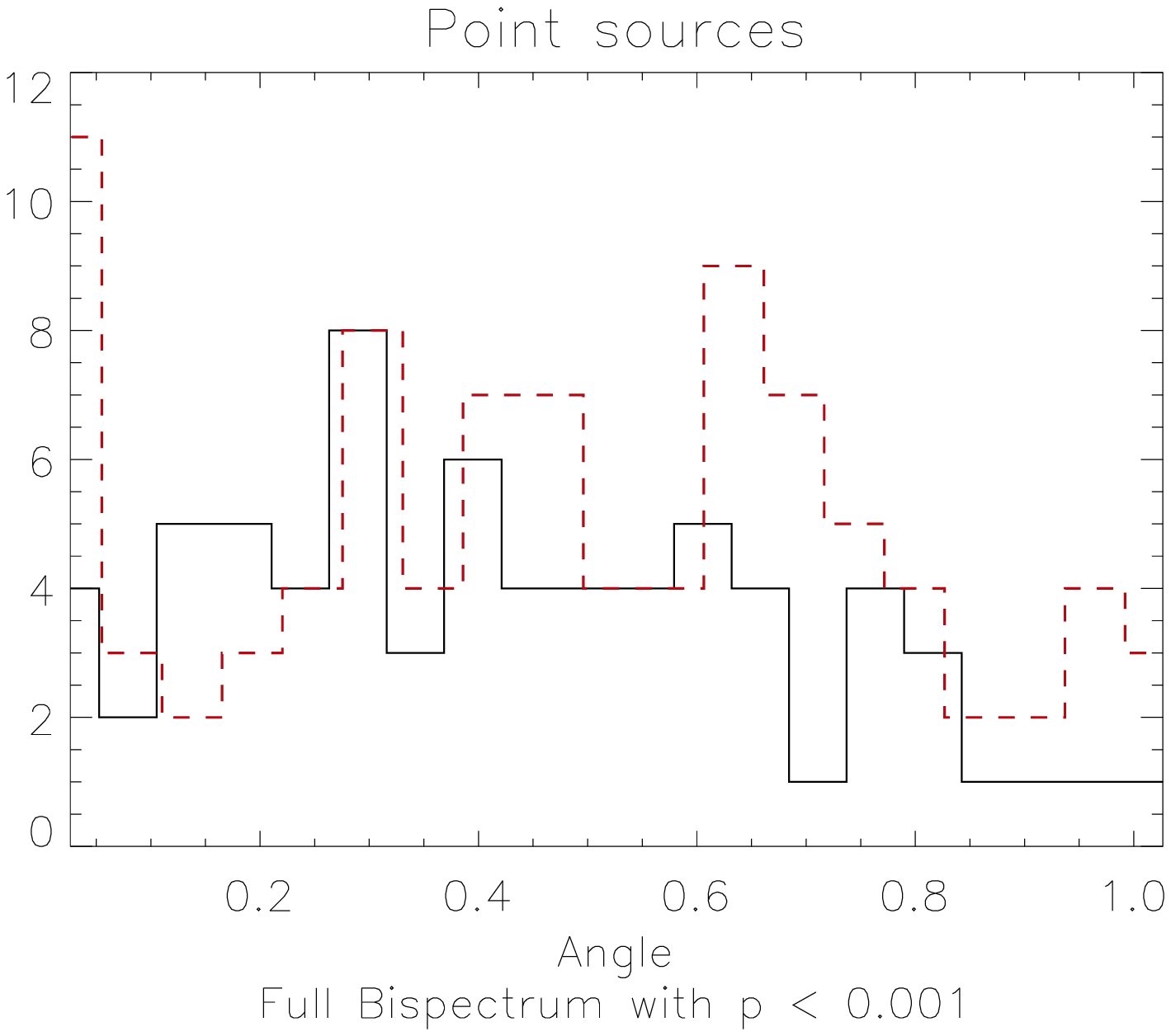}
\includegraphics[width=\columnwidth,height=7cm]{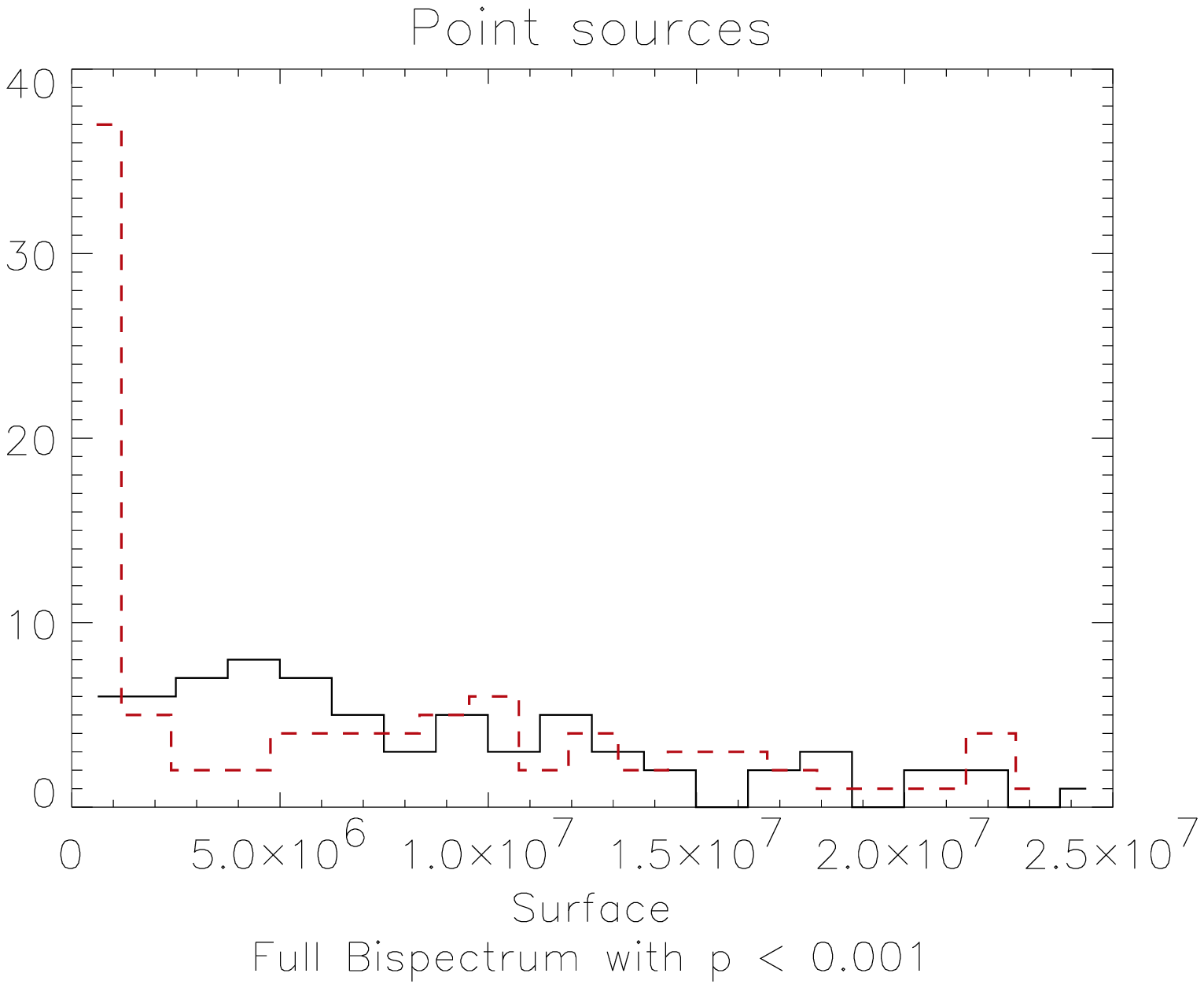}
\end{center}
\caption{For the point sources: The KS probabilities lower than $10^{-3}$. 
In the
surface-angle plane (upper panel) defined by the triplets, the asterisks
are for the comparison Gaussian counterparts versus reference set, and the
diamonds for the comparison non-Gaussian maps versus reference set. The 
middle panel shows the distribution of angles for the above selected KS
probabilities. The lower panel shows the distribution of surfaces for the 
same selection. The dashed line is for non-Gaussian versus reference set, and
the solid line for Gaussian versus reference set.}
\label{fig:sa-bifpt}
\end{figure}

Let us now concentrate
onto the KS probabilities that are lower than $10^{-3}$. We display in the
surface-angle plane the results of the non-Gaussian vs reference set
comparison (Fig. \ref{fig:sa-bifpt}, diamonds) and Gaussian counterparts 
vs reference set 
(Fig. \ref{fig:sa-bifpt}, asterisks). We notice that for small surfaces 
(in $\ell$ space, i.e. Fourier space) the majority of points are associated
with the non-Gaussian maps. There are no detections associated with the 
Gaussian vs reference set comparison. This is
even more obvious when we plot the distribution of surfaces for KS
probabilities $<10^{-3}$ (Fig. \ref{fig:sa-bifpt} lower panel, histogram 
of surfaces). The
first bin is indeed dominated by the detections from the non-Gaussian versus
reference set comparison.
The small surfaces in Fourier space are associated with large surfaces
in real space ($x\propto 1/\ell$). The result we obtain indicates that we are 
actually detecting
the non-Gaussian signatures associated with the higher amplitude spots in the
maps (Fig. \ref{fig:maps}, upper left panel) that are also spatially more
extended and lying on a crowded background of smaller sources. The 
histogram of angles (Fig. \ref{fig:sa-bifpt}, middle panel) in turn does not 
show any highly significant difference between the non-Gaussian versus 
reference and Gaussian counterpart versus reference.

\subsubsection{Filaments}
\begin{figure}
\begin{center}
\includegraphics[width=\columnwidth]{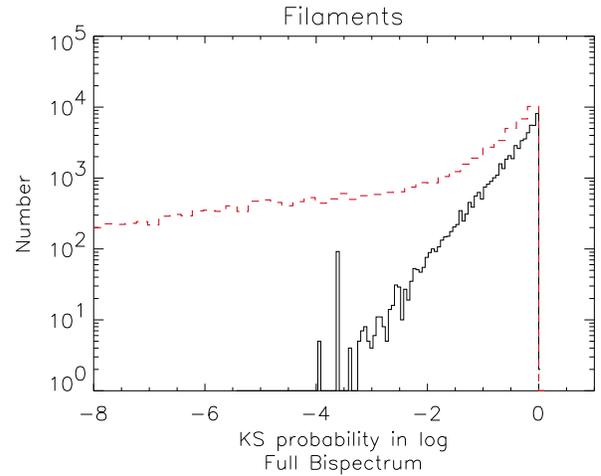}
\end{center}
\caption{For the filaments: Distribution of the KS probabilities for 
the full bispectrum. The dashed line represents the results by comparing the 
non-Gaussian 
maps to the Gaussian reference set. The solid line is for the comparison
Gaussian counterparts versus reference set.}
\label{fig:dist-biir}
\end{figure}

Our last non-Gaussian data set is constituted of the 28 modified IRAS maps
representing the filamentary structures. This process is by far the most
non-Gaussian. The KS test between the distribution of bispectrum
values for the 28 non-Gaussian maps and 99 Gaussian reference maps goes down
to $10^{-20}$ for the full bispectrum.
Similarly to the other processes, we also compare 28
Gaussian realisations to the reference set in order to
estimate the significance of the detections. In the case of the filaments,
there is no doubt that the non-Gaussian signatures are detected. This is
exhibited (Fig. \ref{fig:dist-biir}) by the difference in the distribution 
of KS probabilities obtained when comparing the non-Gaussian maps to the 
Gaussian reference set (dashed line), and the probabilities resulting from the
comparison of 28 Gaussian realisations versus the Gaussian reference set
(solid line).
Unsurprisingly, the {\em meta-statistics} returns a probability of 0 
confirming the very high significance of the non-Gaussian detection. \par
\begin{figure}
\begin{center}
\includegraphics[width=\columnwidth]{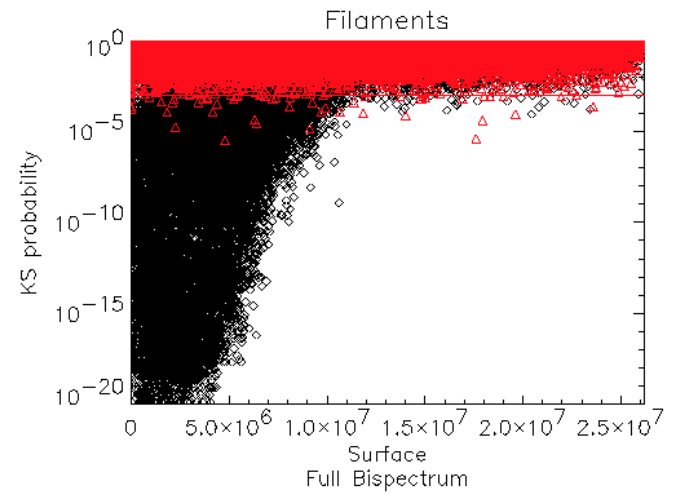}
\end{center}
\caption{The filament maps: KS probabilities 
from the comparison of the full bispectrum as a function of the surface 
defined by the triplet. Black diamonds represent the
results obtained from the comparison non-Gaussian maps versus Gaussian 
reference set. Gray triangles represent the results obtained from the 
comparison Gaussian counterparts versus reference set. The solid 
horizontal line indicates the limit for $P_{KS}=10^{-3}$.}
\label{fig:psbifi-pldira}
\end{figure}

In Fig. \ref{fig:psbifi-pldira}, we display both the Gaussian vs Gaussian
(gray triangles), and non-Gaussian vs Gaussian (black diamonds) points. 
We note that
most of the probabilities in the Gaussian vs Gaussian case lie above the
$10^{-3}$ limit (horizontal solid line). The non-Gaussian character is very 
strongly localised in the triplets with small surfaces. The distribution
as a function of angle is much more uniform.

\begin{figure}
\begin{center}
\includegraphics[width=\columnwidth,height=6.7cm]{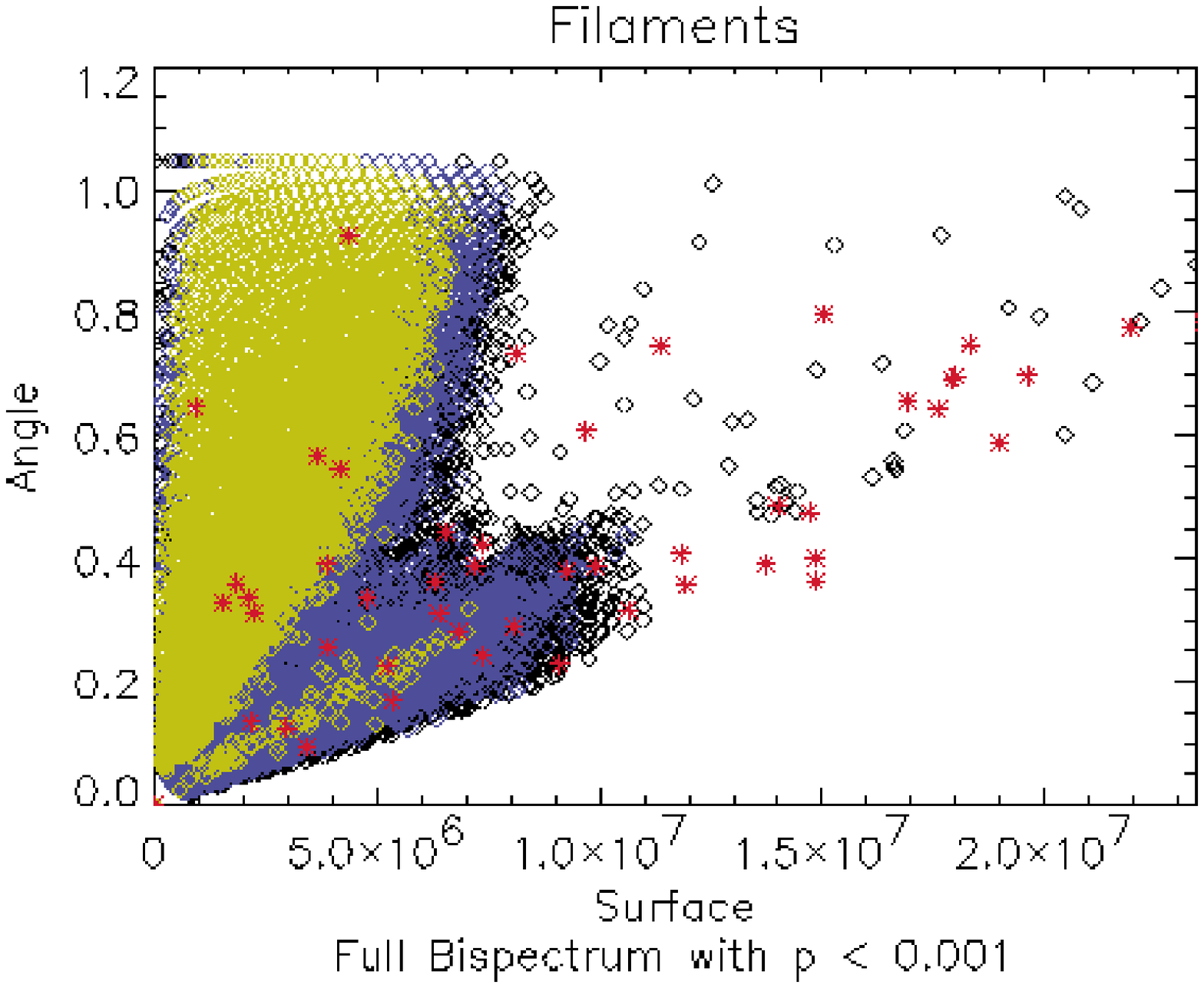}
\includegraphics[width=\columnwidth,height=6.7cm]{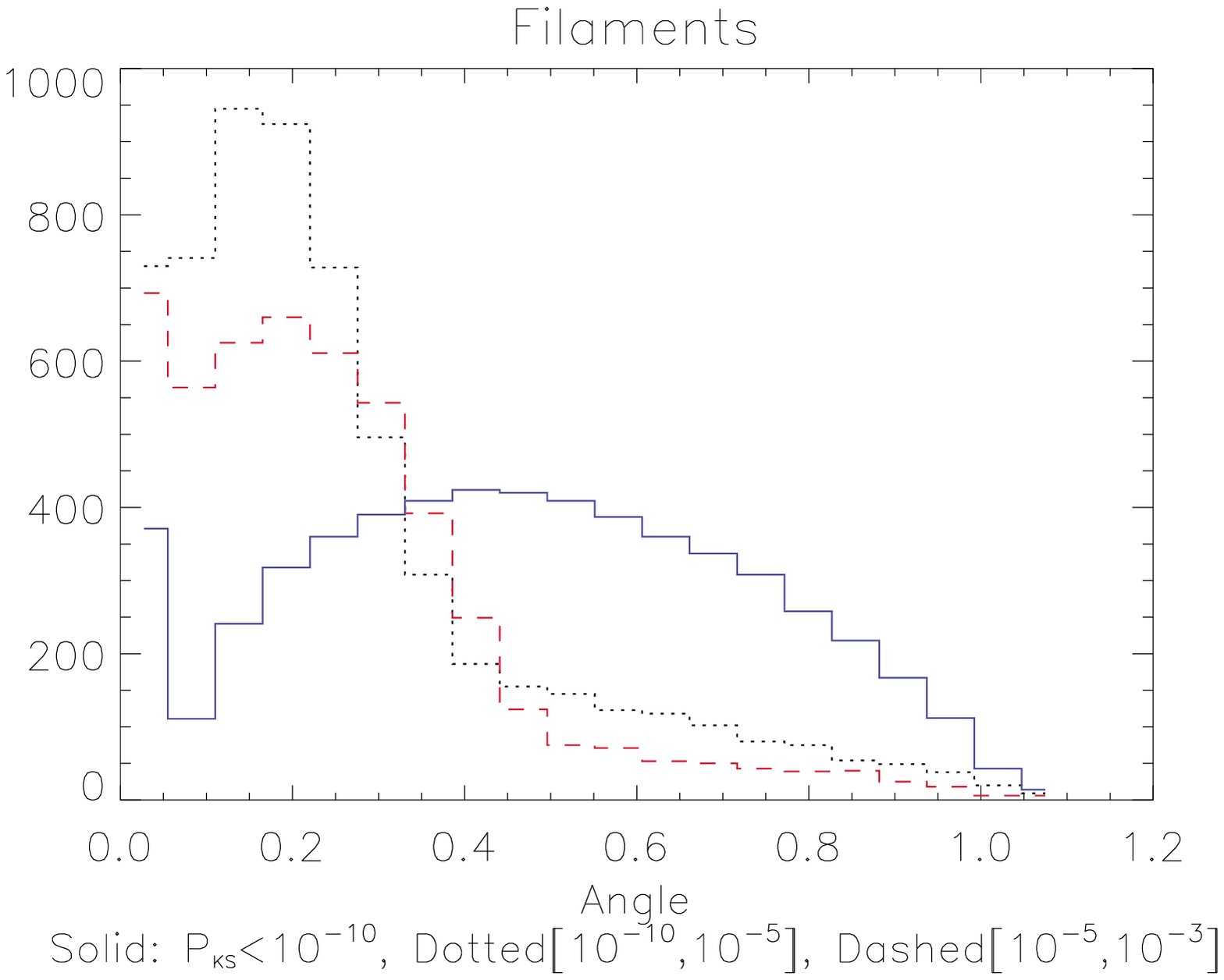}
\includegraphics[width=\columnwidth,height=6.7cm]{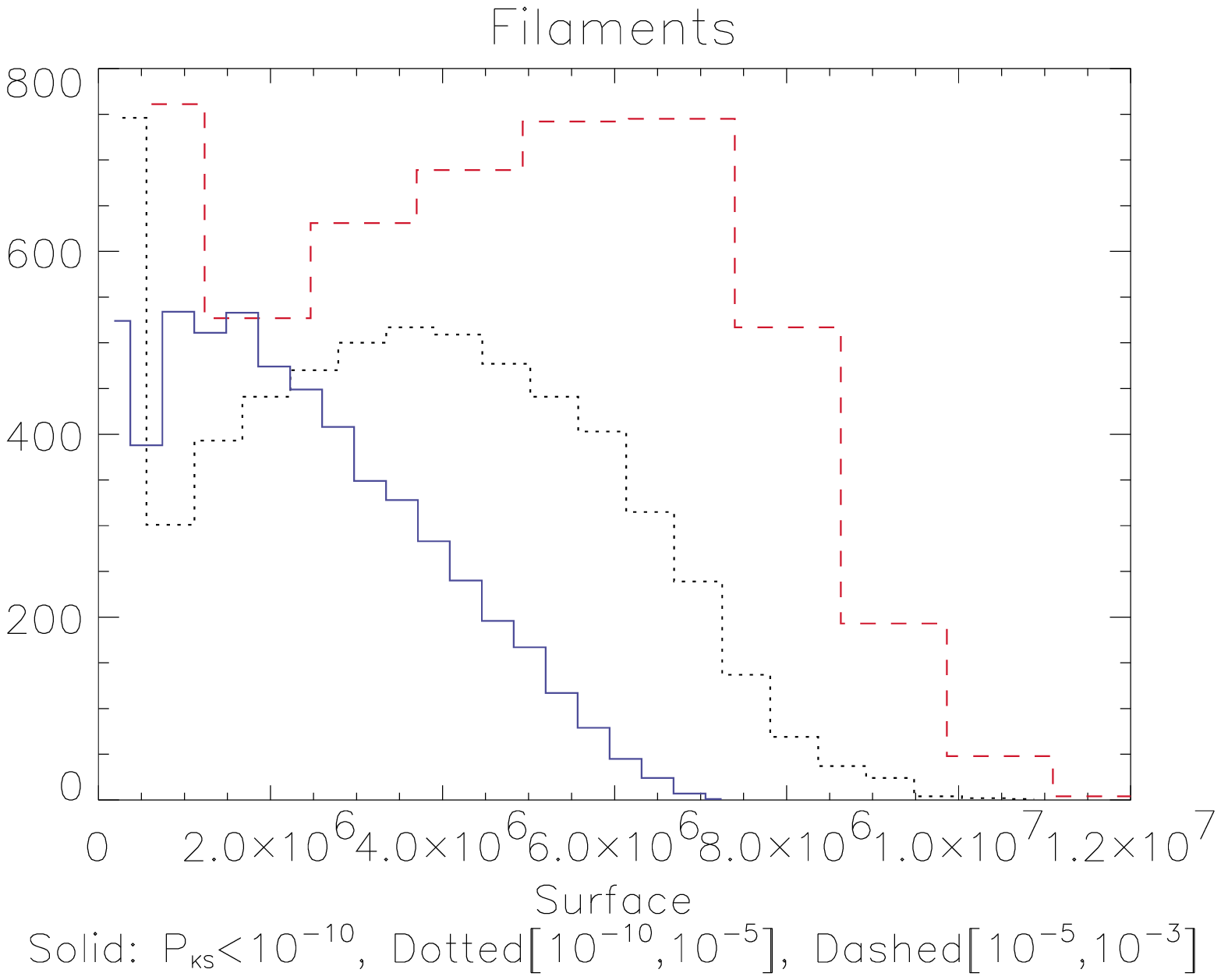}
\end{center}
\caption{For the full bispectrum of the filaments: Distribution of KS 
probabilities in the surface-angle plane defined by the triplets (upper 
panel). Diamonds are for the comparison non-Gaussian maps versus 
Gaussian reference set. Black diamonds are for the KS probabilities 
comprised between $10^{-5}$ and $10^{-3}$, gray diamonds represent 
$10^{-10}<P_{KS}<10^{-5}$, and light gray diamonds stand for 
$P_{KS}<10^{-10}$. The asterisks are for the comparison Gaussian counterparts 
versus reference set. The 
middle panel shows the distribution of angles for the above selected KS
probabilities. The lower panel shows the distribution of surfaces for the 
same selection. Dashed lines are for $10^{-5}<P_{KS}<10^{-3}$, dotted lines
represent $10^{-10}<P_{KS}<10^{-5}$, and solid lines stand for 
$P_{KS}<10^{-10}$.}
\label{fig:sa-bifi}
\end{figure}

If we study the KS probability as a function of both angle and surface
(see figure \ref{fig:sa-bifi}),
we find that the non-Gaussian signatures are mainly localised in two distinct
``plumes''. The first contains triplets with angles basically 
no larger than 0.4 radian
and surfaces that are as large as $10^7$, while the second, larger plume,
covers all angles but only triplets with a slightly smaller surface.
The two histograms, in Fig.~\ref{fig:sa-bifi}, show the number of
triplets within a range of KS probability as a function of either 
surface or smallest angle. They show that the very lowest probabilities
(solid line) are associated with the second, larger plume. They are
distributed over a wide range of angles, but have all a small surface. 
The intermediate probabilities are concentrated in triplets with small
angles but a wider range of surfaces. 
These figures show that the bispectrum allows very
detailed analysis of the non-Gaussianity present in a data set,
if it is detected at a sufficiently high level.
\subsection{Results from the trispectrum}        

Following the same approach as for the bispectrum 
and as explained in detail in section \ref{sec:stat},
we compare for each pair $(\ell,a)$ the distribution 
of the trispectrum values (defined as in 
section \ref{sec:fourier}) for the 
non-Gaussian set and for its Gaussian reference set. 
We compare the sensitivity of the
two estimators $T^{(0)}$ and $T^{(+)}$ (respectively diagonal and 
near-diagonal trispectra) in terms of non-Gaussian detection on 
the point source 
maps since the signal is non-Gaussian but not as extreme as in the 
filaments.

As a significance discriminator, we use the equivalent 
results from the comparison between the set of Gaussian 
counterpart maps and the Gaussian reference set.
We checked that the normalisation of the trispectrum to the power 
spectrum does not affect the resulting KS probabilities significantly.
\subsubsection{$\chi^2$ maps}
\begin{figure}
\begin{center}
\includegraphics[width=\columnwidth]{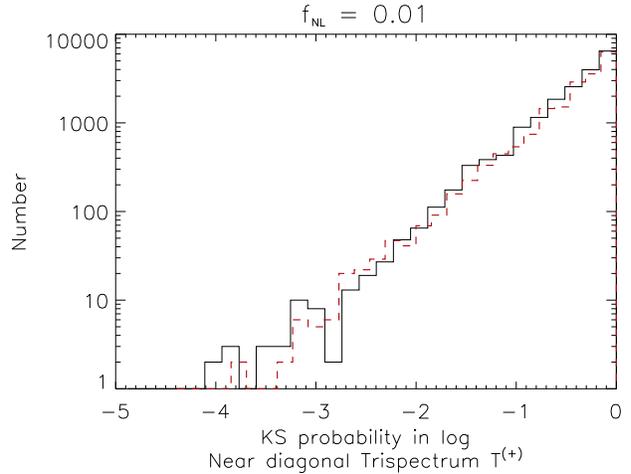}
\end{center}
\caption{For the $\chi^2$ maps with non-linear coupling factor $f_{NL}=0.01$: 
Distribution of the KS probabilities obtained by i) comparing the 
trispectrum estimators of the non-Gaussian maps to those of the Gaussian 
reference set (dashed line), and ii) comparing the trispectrum estimators of 
the Gaussian counterparts to those of the Gaussian reference set (solid line).}
\label{fig:dis-trichi}
\end{figure}

For the $f_{\mathrm NL}= 0.01$ $\chi^2$ maps and Gaussian maps, the
smallest KS probabilities are a few times $10^{-5}$, see
Fig.~\ref{fig:dis-trichi}. The {\em meta-statistics} confirms that
we do not see any significant deviation from the Gaussian hypothesis: The 
orthonormalised
$T^{(+)}$ estimator leads to a {\em meta--}$P_{KS} = 0.13$ for a non-linear
coupling constant of $0.01$.

It is not surprising that the trispectrum does not detect any
non-Gaussian signature. Just as for the excess kurtosis of the 
pixel distribution
in real space (see section \ref{sec:res-cpf}), the lowest order
contribution to the trispectrum is proportional to $f_{NL}^2$.
This makes it naturally much less suited to find this kind of non-Gaussian
signals. This fact can however be used as a counter-check, since if we
believe to have detected a $\chi^2$ type signal in the bispectrum,
we should not find it when applying the trispectrum to perfectly clean 
data.
\subsubsection{Point sources}
The histogram of probability values in figure \ref{fig:dis-tript} (upper 
panel) does
not show any extreme differences for $T^{(+)}$ between the Gaussian vs 
Gaussian and the non-Gaussian vs Gaussian comparisons, with only one value
at  $(\ell,a)\simeq(300,0)$ having a particularly low probability
of $10^{-8}$. The {\em global
\begin{figure}
\begin{center}
\includegraphics[width=\columnwidth]{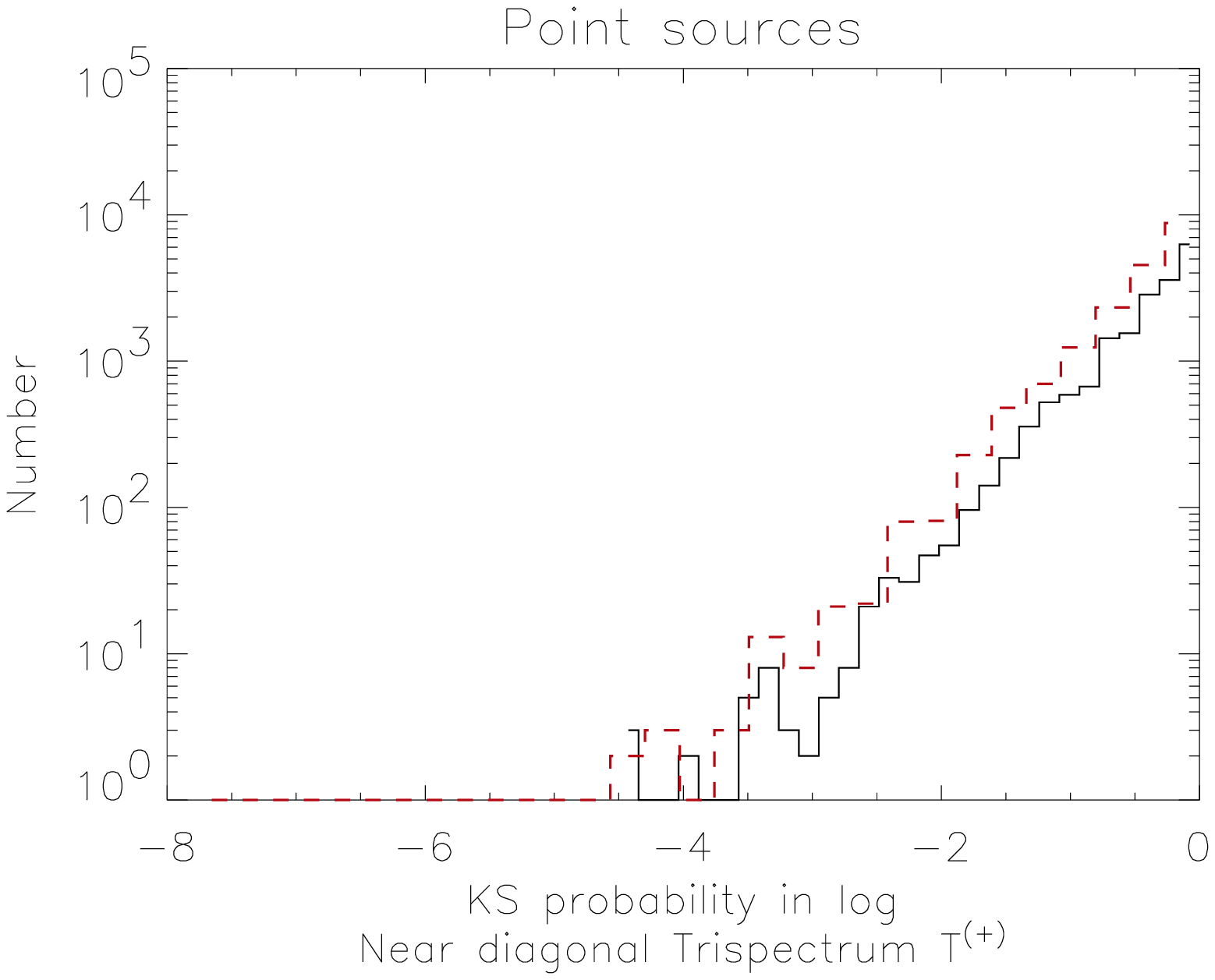}
\includegraphics[width=\columnwidth]{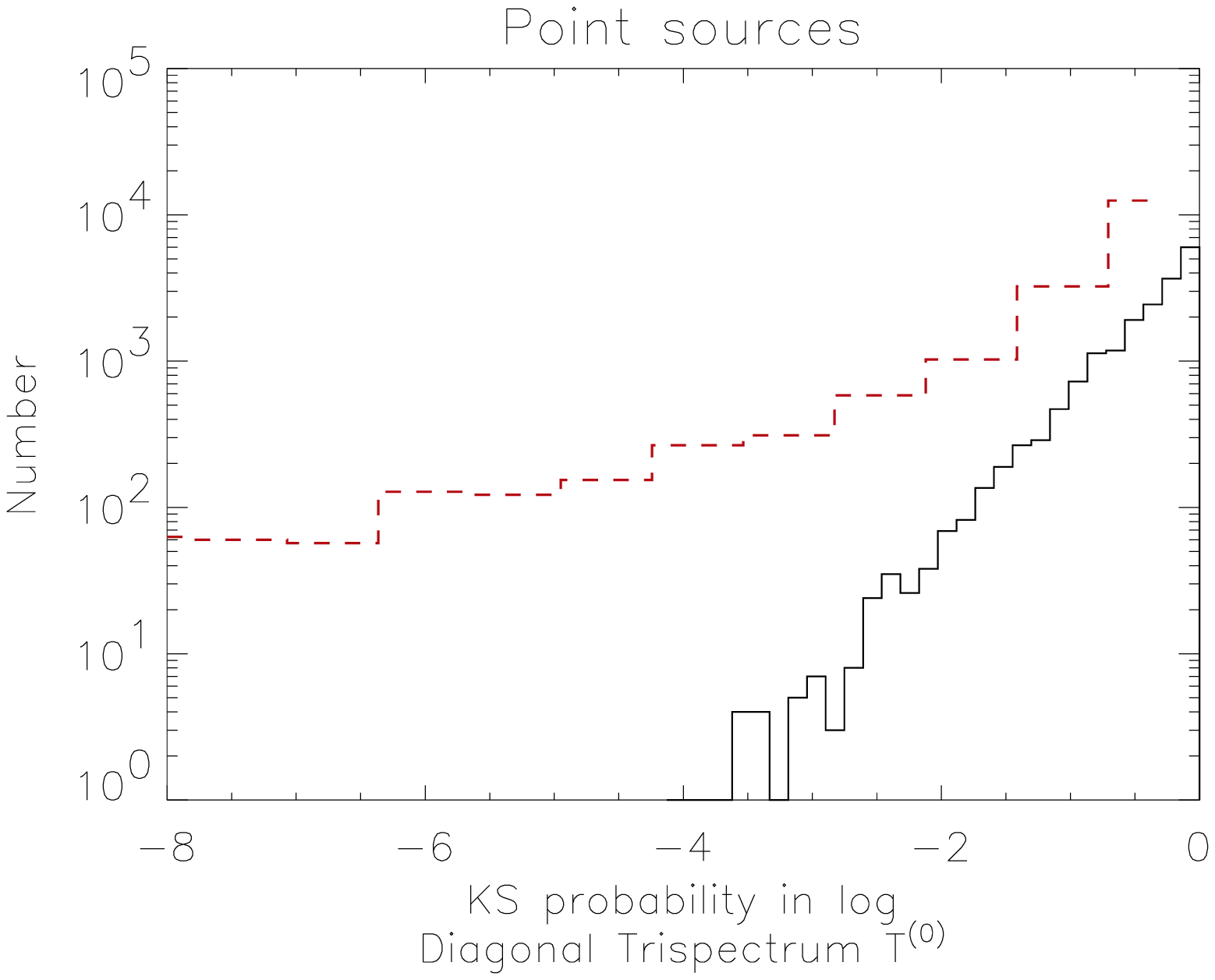}
\end{center}
\caption{For the point sources: Distribution of the KS probabilities 
obtained by i) comparing the 
trispectrum estimators of the non-Gaussian maps to those of the Gaussian 
reference set (dashed line), and ii) comparing the trispectrum estimators of 
the Gaussian counterparts to those of the Gaussian reference set (solid line).
The upper panel is for the near-diagonal trispectrum ($T^{(+)}$) and the lower 
panel for the diagonal trispectrum estimator ($T^{(0)}$).}
\label{fig:dis-tript}
\end{figure}
probability} for being Gaussian is $8.8\,10^{-4}$. This changes dramatically
if we look at $T^{(0)}$ (Fig. \ref{fig:dis-tript}, lower panel). The 
diagonal estimator detects a very clear
non-Gaussian signal, and the {\em meta-statistics} returns a probability
of $0$. It is quite surprising that the two estimators show such a
huge difference, although they are indeed completely independent,
since $T^{(+)}$ does not have any Gaussian contribution, as opposed
to $T^{(0)}$.
\begin{figure}
\begin{center}
\includegraphics[width=\columnwidth]{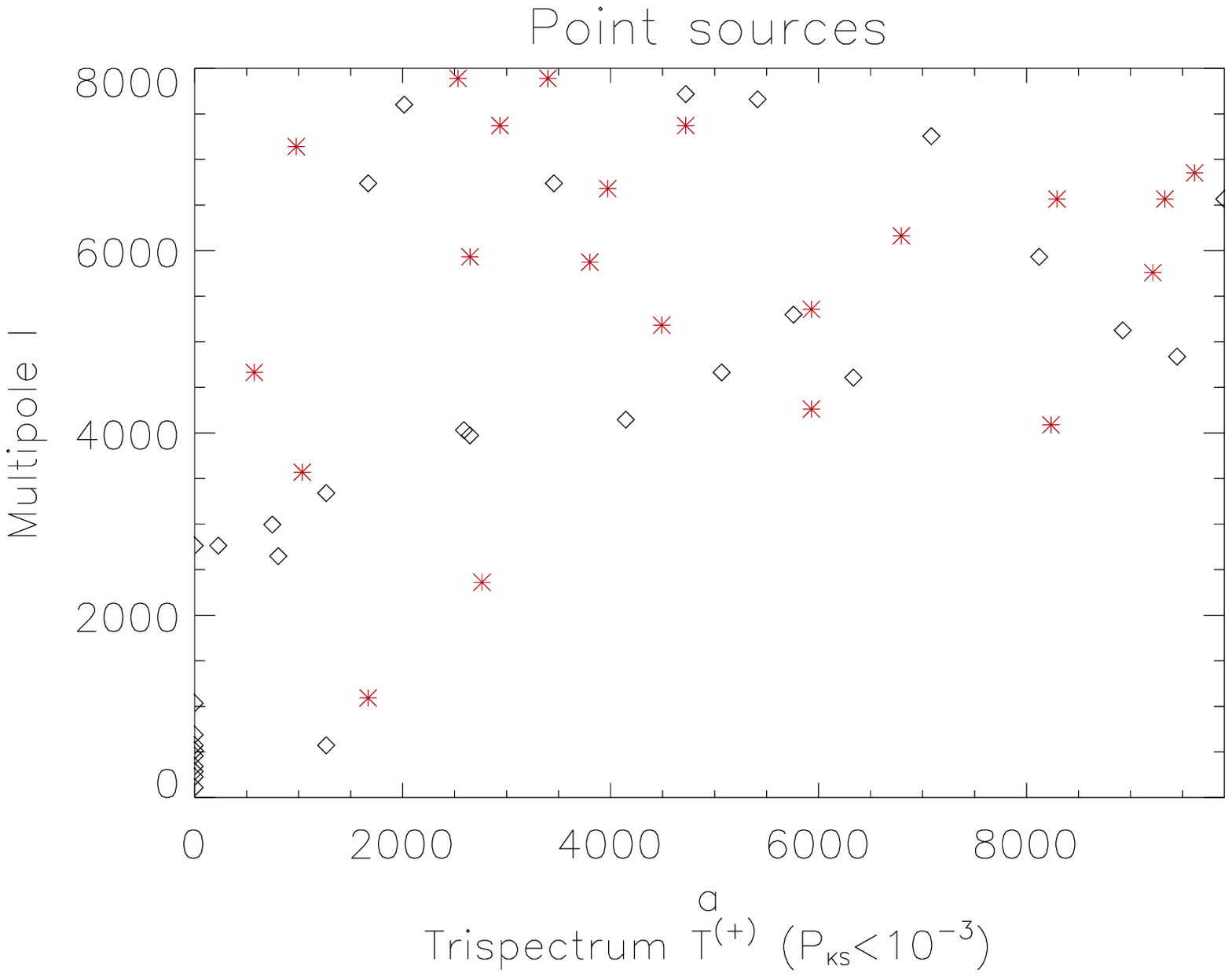}
\includegraphics[width=\columnwidth]{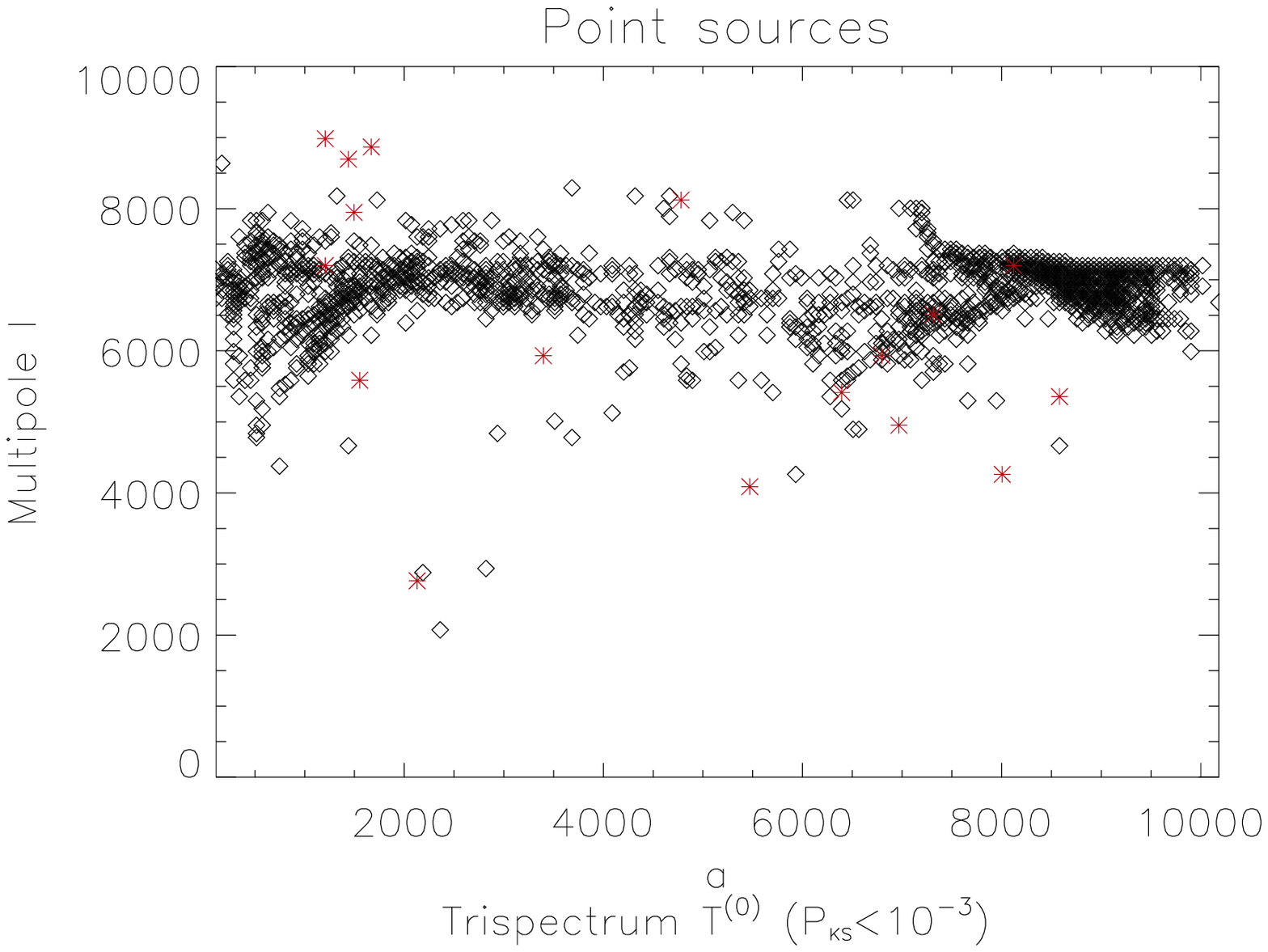}
\end{center}
\caption{For the trispectrum of the point sources: Distribution, in 
($\ell$,$a$) space, of the 
KS probabilities lower than $10^{-3}$. Asterisks are for the comparison
between Gaussian counterparts and Gaussian reference set. Diamonds stand
for the non-Gaussian versus Gaussian reference set comparison. 
The upper panel is for the near-diagonal trispectrum. Most diamonds are 
found at very small $a$ in this case. The lower panel is for the diagonal 
trispectrum estimator.}
\label{fig:la-tript}
\end{figure}

Looking at the distribution of the low-probability points in $(\ell,a)$
space (Fig. \ref{fig:la-tript} upper and lower panels), we find that 
$T^{(0)}$ is localised around $\ell \approx 7000$ 
and seems to prefer either very small or very large values of $a$,
corresponding to oblong configurations. The weak signal of $T^{(+)}$ 
seems to be mainly due to points with low $\ell$ and very small $a$
(diamonds in Fig. \ref{fig:la-tript} upper panel). 
\subsubsection{Filaments}
\begin{figure}
\begin{center}
\includegraphics[width=\columnwidth]{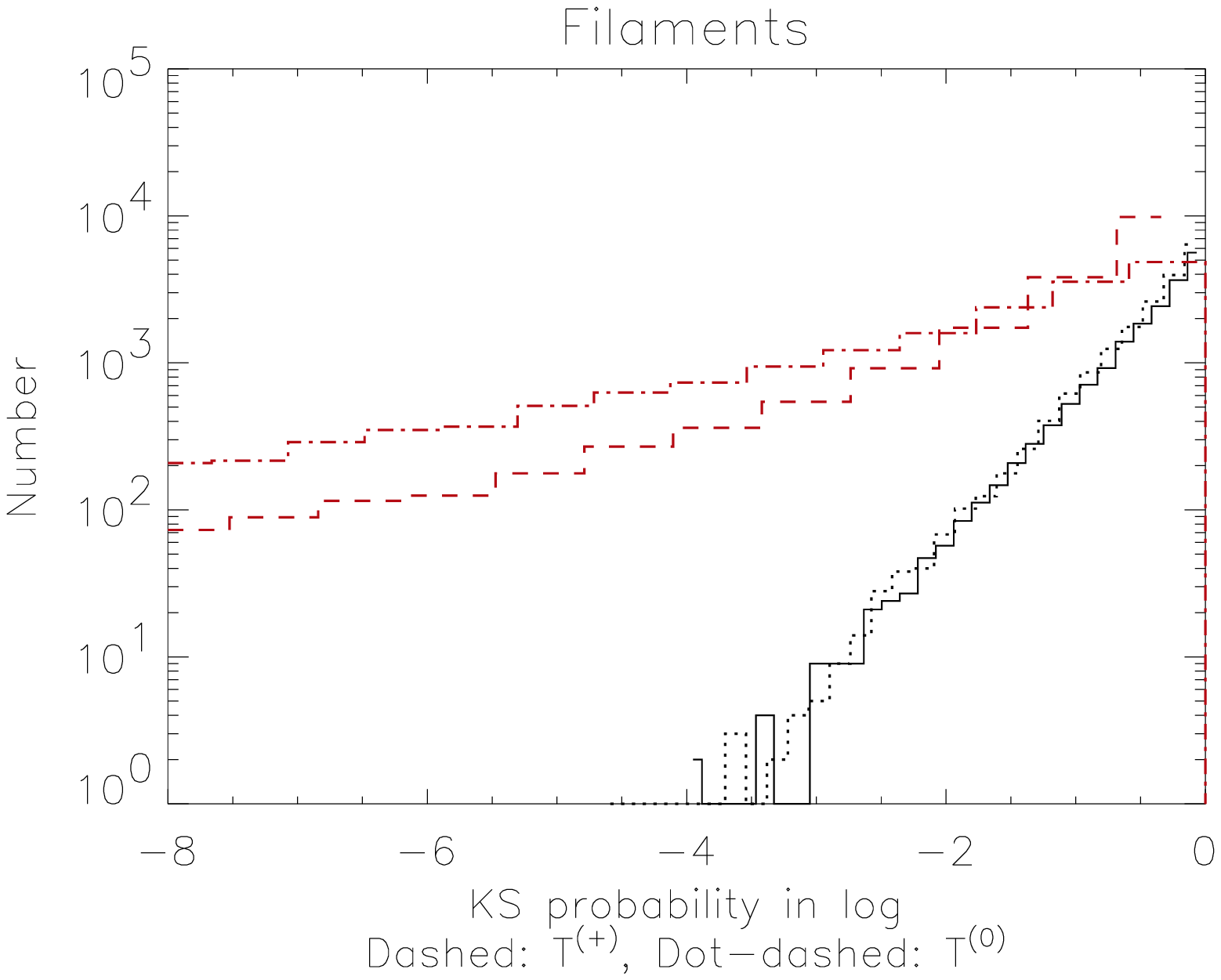}
\end{center}
\caption{For the trispectrum of the filaments: Distribution of KS 
probabilities.
The solid line is for the comparison of Gaussian counterparts vs Gaussian 
reference set for the near-diagonal estimator ($T^{(+)}$). The dashed line
is for the comparison of the non-Gaussian maps to reference set for the
near-diagonal trispectrum. The dotted line is for the comparison of Gaussian 
counterparts vs reference set for the diagonal trispectrum ($T^{(0)}$). 
The dot-dashed line is for the non-Gaussian maps vs reference set for the
diagonal estimator.}
\label{fig:dis-trira}
\end{figure}

For the case of the non-Gaussian filament maps,
the KS test gives probabilities that the 
non-Gaussian signal is issued from the same process 
as the Gaussian reference set as low as a few $10^{-20}$. 
Figure \ref{fig:dis-trira} shows the difference in the distributions 
of the KS probabilities for diagonal and near-diagonal trispectrum. 
Furthermore, 
the {\em meta-probability} obtained by applying the KS test on these 
probability distributions returns 0. In this case and similarly to the 
bispectrum, the non-Gaussian signatures are consequently
undoubtedly detected. However, again the diagonal estimator seems more
sensitive than the near-diagonal trispectrum. As shown in Fig. 
\ref{fig:dis-trira}, there are many more (twice as many) low probabilities 
(high confidence detection) in the diagonal case than in the near-diagonal.
\par

\begin{figure}
\begin{center}
\includegraphics[width=\columnwidth,,height=6.7cm]{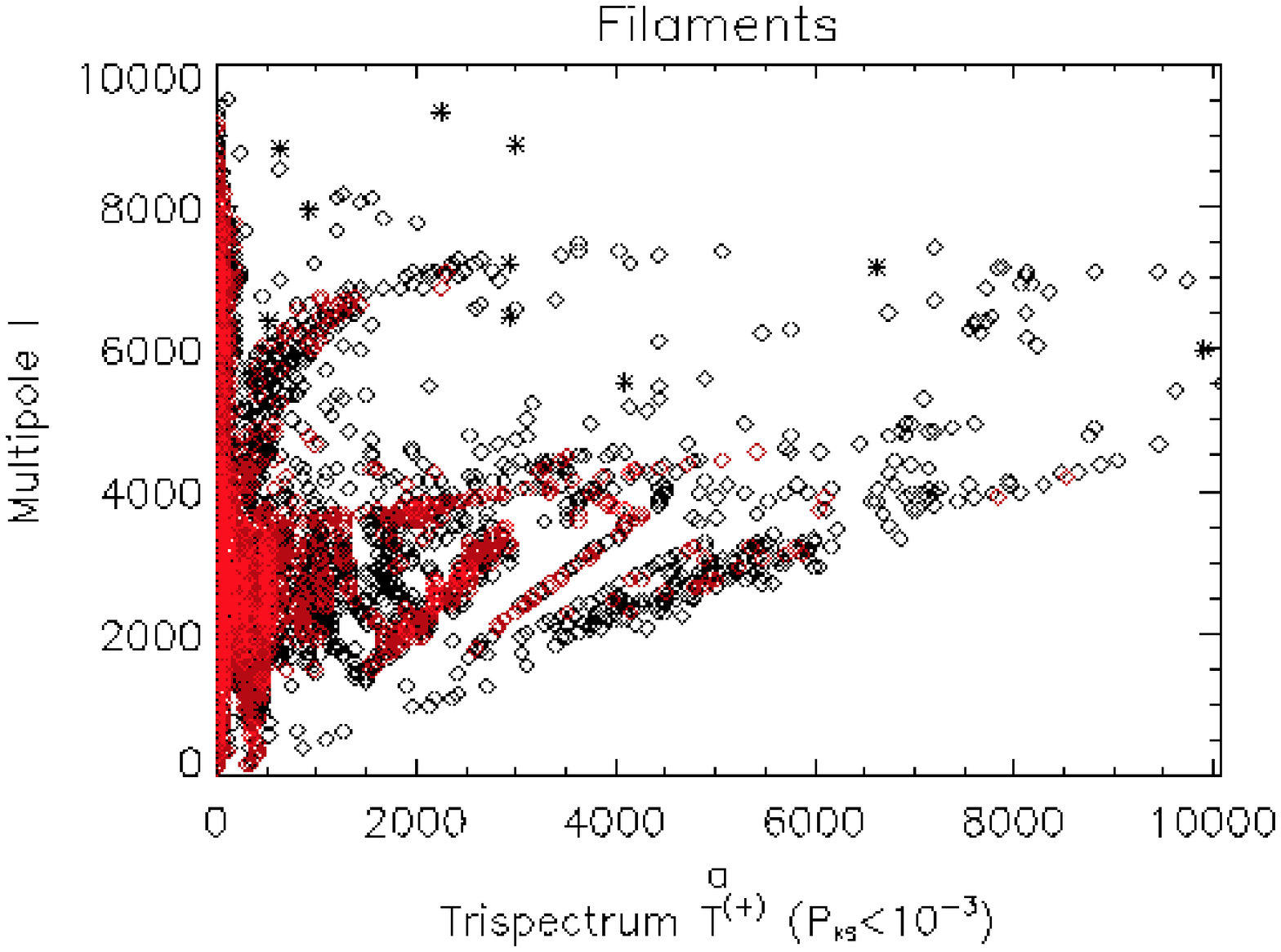}
\end{center}
\begin{center}
\includegraphics[width=\columnwidth,height=6.7cm]{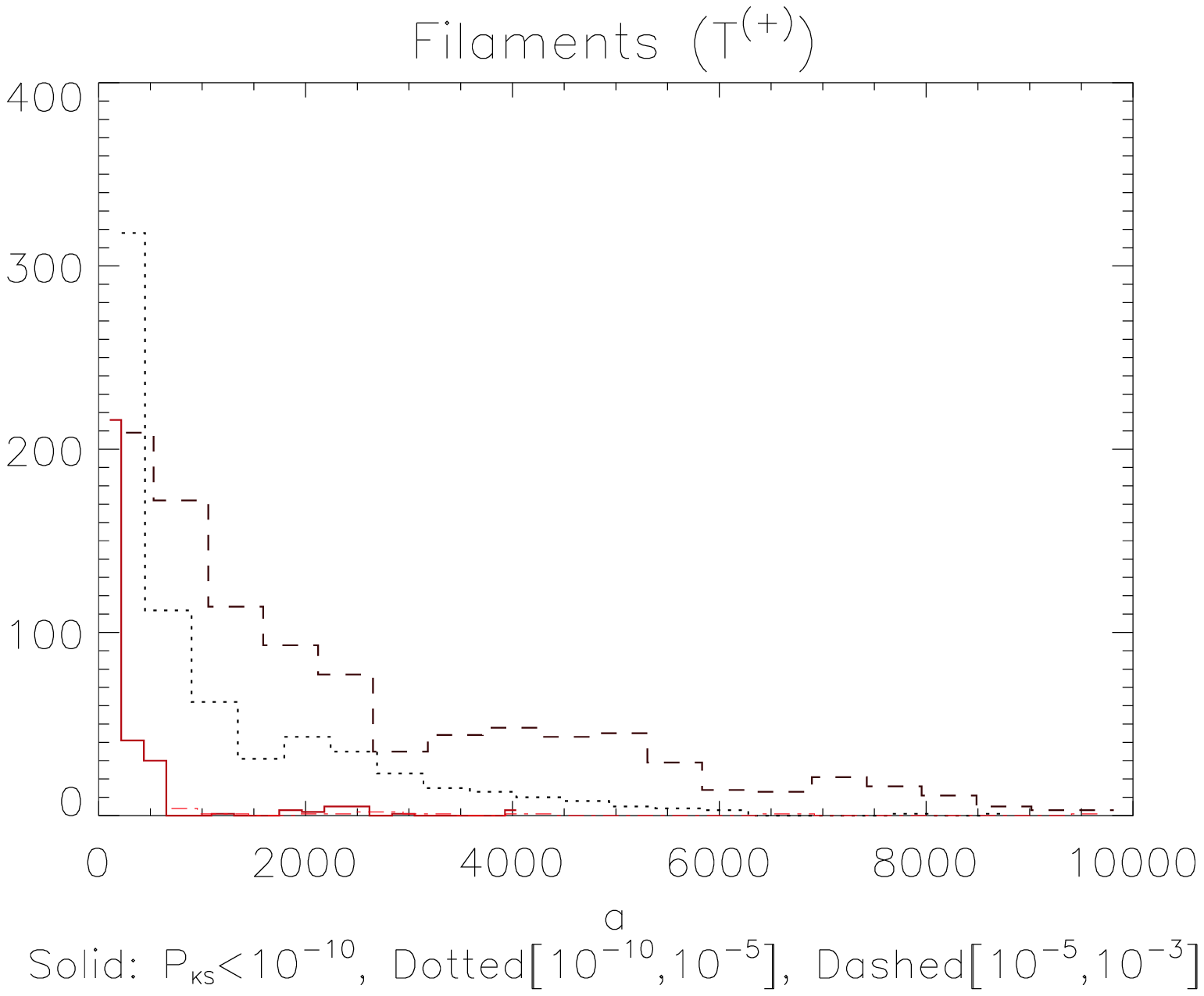}
\end{center}
\begin{center}
\includegraphics[width=\columnwidth,height=6.7cm]{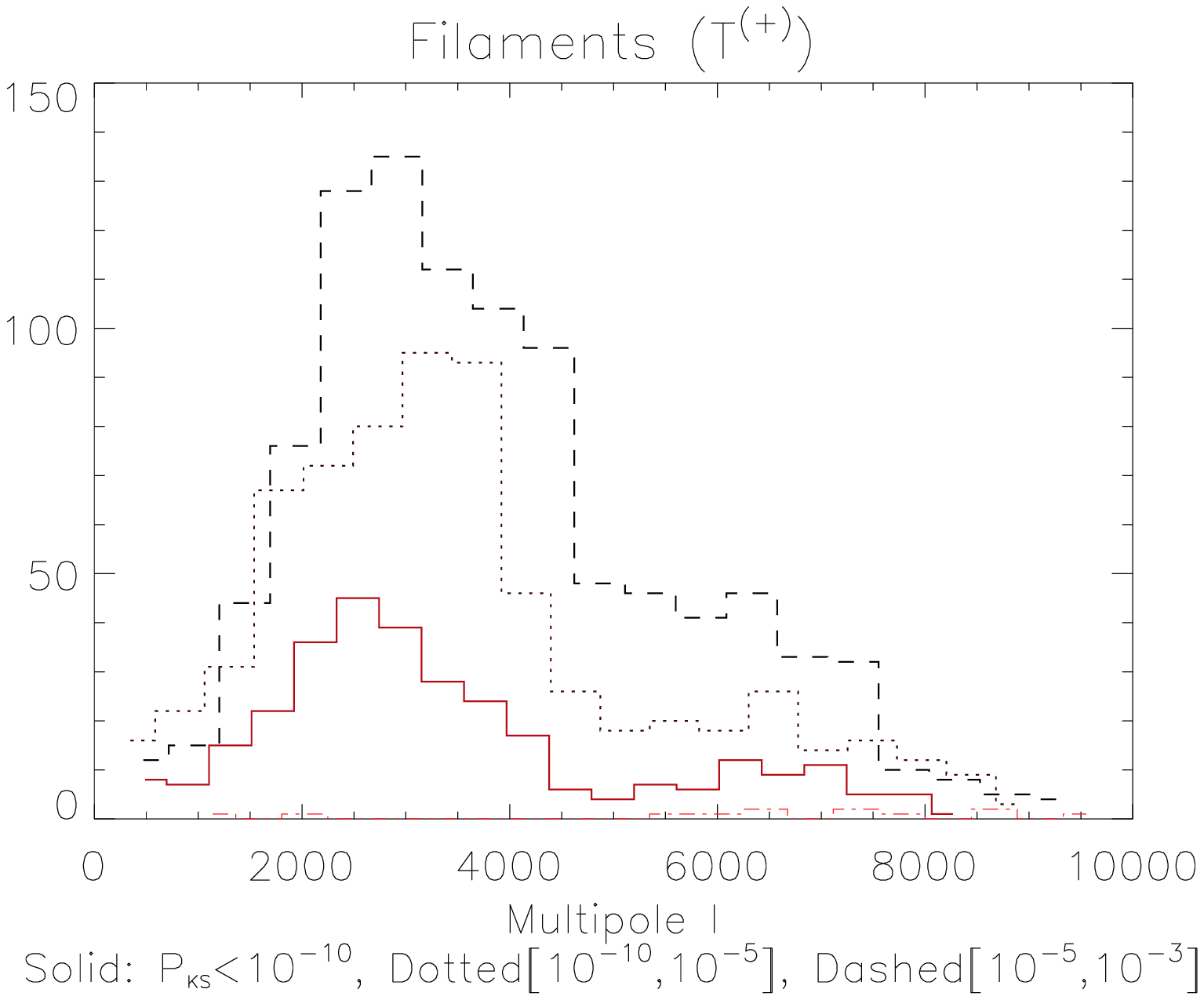}
\end{center}
\caption{For the near-diagonal trispectrum estimator of the filaments: 
The KS probabilities lower than $10^{-3}$ in the
($\ell$,$a$) space (upper panel). The asterisks
are for the comparison Gaussian counterparts versus reference set, and the
diamonds for the comparison non-Gaussian maps versus reference set. The 
middle panel shows the distribution of $a$ for the above selected KS
probabilities. The lower panel shows the distribution of $\ell$ for the 
same selection. Dashed lines are for $10^{-5}<P_{KS}<10^{-3}$, dotted lines
represent $10^{-10}<P_{KS}<10^{-5}$, and solid lines stand for 
$P_{KS}<10^{-10}$.}
\label{fig:la-trira-tl2}
\end{figure}

\begin{figure}
\begin{center}
\includegraphics[width=\columnwidth,height=6.7cm]{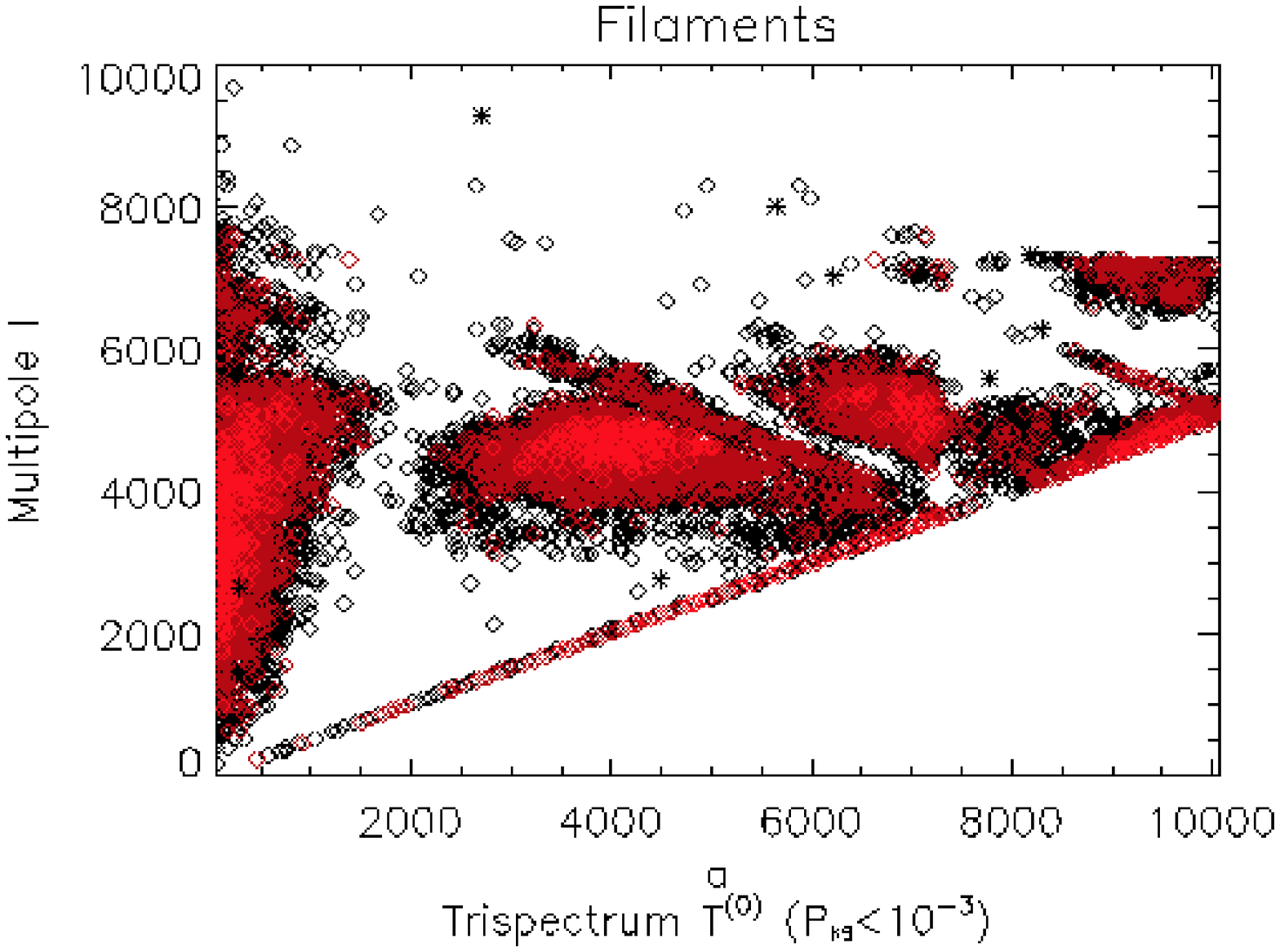}
\end{center}
\begin{center}
\includegraphics[width=\columnwidth,height=6.7cm]{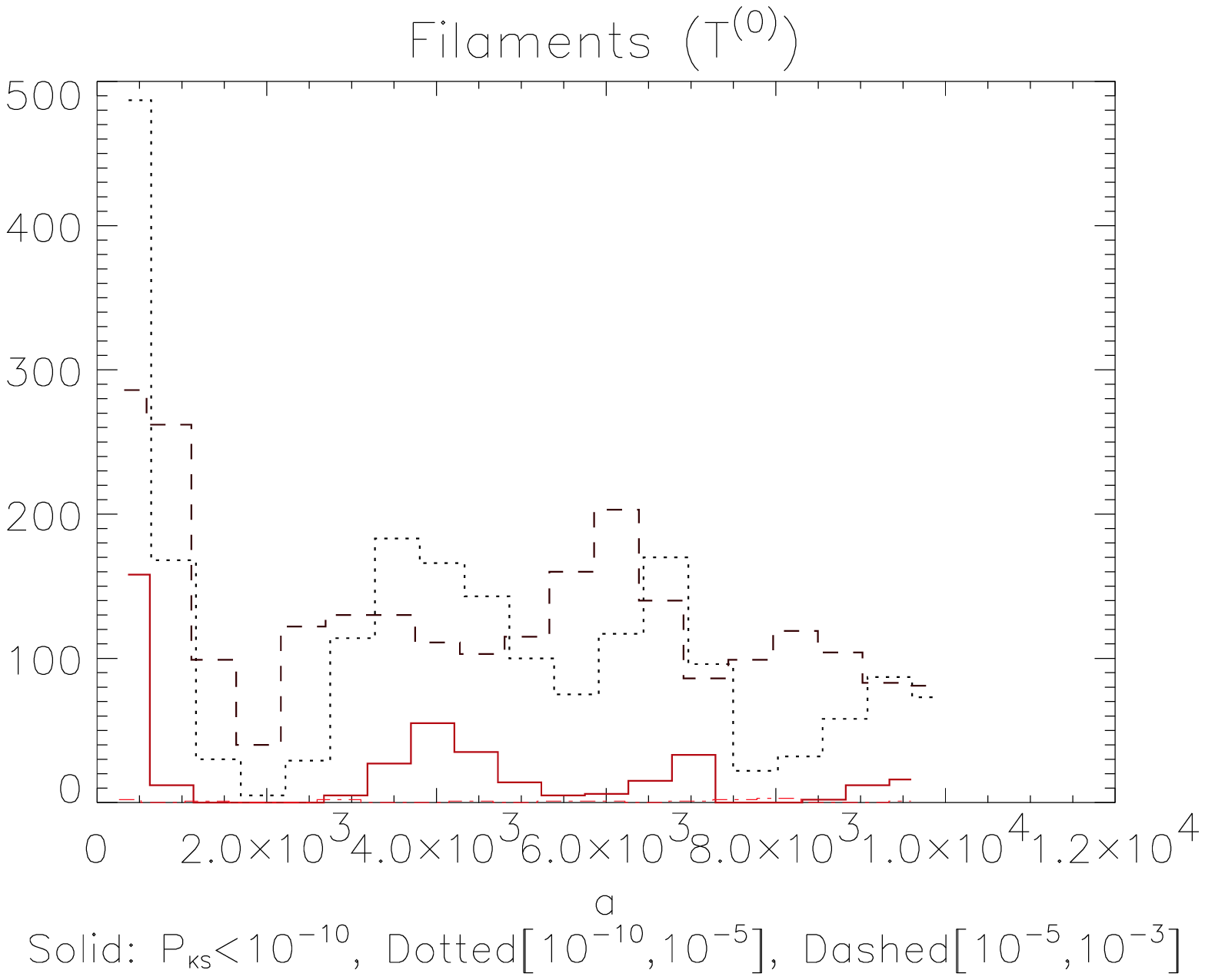}
\end{center}
\begin{center}
\includegraphics[width=\columnwidth,height=6.7cm]{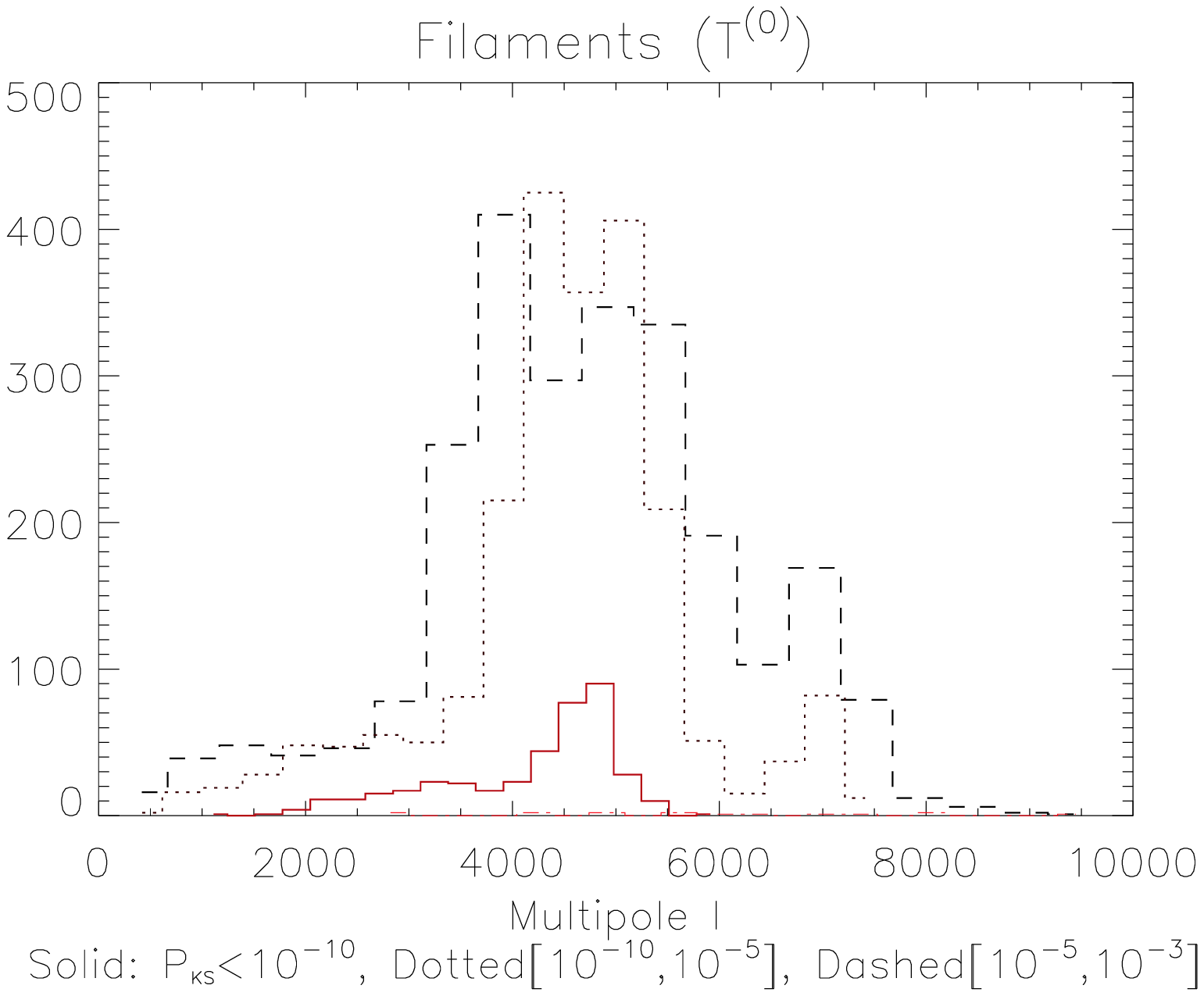}
\end{center}
\caption{For the diagonal trispectrum estimator of the filaments: The KS 
probabilities lower than $10^{-3}$ in the
($\ell$,$a$) space (upper panel). The asterisks
are for the comparison Gaussian counterparts versus reference set, and the
diamonds for the comparison non-Gaussian maps versus reference set. The 
middle panel shows the distribution of $a$ for the above selected KS
probabilities. The lower panel shows the distribution of $\ell$ for the 
same selection. Dashed lines are for $10^{-5}<P_{KS}<10^{-3}$, dotted lines
represent $10^{-10}<P_{KS}<10^{-5}$, and solid lines stand for 
$P_{KS}<10^{-10}$.}
\label{fig:la-trira-tldiag}
\end{figure}

We try to visualise the morphological information contained in the
trispectrum coefficients by displaying in $(\ell,a)$ space 
the KS probabilities (Figs. \ref{fig:la-trira-tl2}
and \ref{fig:la-trira-tldiag}, upper panels).
Those associated with the detection of the non-Gaussian signatures are in 
diamonds, and those associated with statistical fluctuations in the Gaussian
realisations are in asterisks. 
We plot in figures \ref{fig:la-trira-tl2}
and \ref{fig:la-trira-tldiag} the distribution of $a$ and $\ell$ in
bins of KS probabilities for the trispectrum ($T^{(+)}$) and ($T^{(0)}$).
Both trispectra prefer small values of $a$ (or the diagonal $a \sim 2\ell$
which correspond to similar configurations), which are associated with strongly
elongated structures. The diagonal trispectrum exhibits several other low-
probability ``blobs'' at intermediate $a$ values, but the corresponding
histogram (the middle panel of Fig. \ref{fig:la-trira-tldiag}) shows that these
are somewhat less important. On the $\ell$ axis, $T^{(0)}$ seems non-Gaussian
mainly for values between $4000$ and $6000$, while $T^{(+)}$ rather detects
a signal around $\ell = 2000 - 4000$. All in all, the two trispectra together
show a rich structure with clear concentrations of non-Gaussian signatures,
not just random scatter.
%
\section{Conclusions}

In the present study, we investigate two of the major families of
non-Gaussian estimators, namely Fourier-space based methods (the bi- and
trispectrum) and wavelet-space based methods (the skewness and excess kurtosis of
the wavelet coefficients). They are applied to three quite
different data sets chosen to represent a rather complete sample
of possible non-Gaussian signatures (namely: point sources, filaments,
non-linearly coupled signals). Additionally, the methods were
applied to two sets of Gaussian realisations (called reference set and
counterpart set) with the same power
spectra as the non-Gaussian maps.  We then use the Kolmogorov-Smirnov 
test to quantify the level of detection of the non-Gaussian signatures 
by comparing the distributions of the estimator values 
for the non-Gaussian maps versus the
Gaussian reference set, and for the Gaussian counterpart set versus
the reference set. The first comparison returns directly an estimate
of the non-Gaussianity detected by a method, while the second case 
serves to illustrate the statistical fluctuations within the Gaussian 
signal itself. Furthermore, we have checked that other
statistical tests such as the Kuiper and Anderson-Darling tests give the
same results.

We find that the filaments represent a highly non-Gaussian signal.
This statistical character is undoubtedly detected by both the Fourier
and wavelet based methods in the three and four point estimators.
For the point source maps, the non-Gaussian signatures are very
significantly detected with the excess kurtosis of the wavelet coefficients
and the diagonal estimator of the trispectrum, but less significantly
with the bispectrum and very marginally with the skewness of the wavelet
coefficients. This is consistent
with the moments of the maps, where the non-Gaussianity shows up in
the excess kurtosis as well. It is also expected from the physical nature of
the signal: The maps contain the same number of negative and positive 
sources, on average, which suppresses the skewness.
The $\chi^2$ type maps show just the inverse
behaviour, in that it is now the three point tests (skewness of the
pixel distribution and of the wavelet coefficients as well as the
bispectrum) which detect the non-Gaussian character best. As mentioned
earlier this is related to the signal itself which is constructed by
adding a positive contribution. 

When assessing the relative level of detection, we notice that
the wavelet analysis finds the non-Gaussian signatures in the $\chi^2$ maps
with a non-linear coupling coefficient of 1\% at a very high confidence
but only at the first decomposition scale (3 arc minute scale),
while the bispectrum detection is somewhere between $3$ and $4 \sigma$.
Wavelets are therefore better at placing constraints on $f_{\rm NL}$,
a result which confirms the conclusion of a recent analysis of the 
COBE-DMR maps with wavelets by \cite{cayon2002}. The point source maps
are clearly recognised as non-Gaussian by all methods. 
Finally we note that
the diagonal trispectrum estimator is surprisingly much more sensitive 
than the nearly diagonal estimator, even though the latter does
not contain a Gaussian contribution. This shows 
that care must be taken if not all components of a given estimator
are computed, as the signal may be extremely localised.
A forthcoming study will address this question in more detail.

Comparing the methods on a more fundamental level, we find
as an additional advantage of the wavelet-based approach that
it allows us to associate the non-Gaussian signatures with the features that
have caused them in the map, at all the scales where they exist. Wavelet 
decomposition thus permits to take benefit of any scale-scale 
correlations that might exist in the non-Gaussian signal. This
can be done directly in the wavelet space or by selecting the coefficients 
and reconstructing the signal with the inverse wavelet transform. These aspects
have not been investigated in the present study. 

However, the wavelet based estimators are lacking one important property
of the Fourier approach. The bi- and 
trispectrum can be analytically predicted without resorting to
Monte Carlo based methods. They can thus be directly 
related to physical phenomena, which
explains their interest for cosmology. Additionally, the bi- and
trispectrum allow us to probe a very large number of geometrical
configurations for the triplets and quadruplets as compared to the
limited number of decomposition scales investigated by the wavelets. 

The present comparison can be
extended in terms of general strategy proposed to the groups analysing 
large CMB data sets. If we try to condense it into a few recommendations, we 
end up with the following steps:
\begin{enumerate}
\item {\em Look at the higher order moments in pixel space and the CPF first.} 
The cumulative probability function is
very easy and fast to calculate. Nonetheless, it seems to have a
similar power as the bi- and trispectrum where just the detection
of non-Gaussian signatures is concerned. It therefore provides a quick
check if one can expect to find a strong signal with the more sophisticated
analysis techniques. We recommend that the maps are deconvolved with
their power spectrum before testing the CPF.
\item {\em Apply wavelet based methods second.} Wavelets have the highest
detection power of the methods presented here, and are only moderately 
more expensive to apply than the CPF test. If indeed a signal is detected,
they can reconstruct its origin in pixel space.
\item {\em Use Fourier methods for a detailed analysis.} Together, the
bi- and trispectrum furnish a large number of coefficients associated 
with different geometrical configurations. As the
number of theoretical predictions for the bi- and trispectrum from various 
sources of non-Gaussian signatures is increasing, this allows a precise
characterisation of any strong and detected non-Gaussianity in a map, and
the use of accurate matched filters to determine its possible origin.
\end{enumerate}
Clearly, combining both wavelet and Fourier based analyses seems the best
strategy for studying the non-Gaussian signals. Together, they allow for 
very significant detections of the non-Gaussian signatures and a large 
amount of spatial information. In
addition, they offer a theoretical basis to relate the physical processes 
to the measured quantities. Therefore, these methods offer currently the
best chances of success when trying to identify the origin of the various
sources of non-Gaussian signatures that are expected to contribute to the 
CMB signal.
\begin{acknowledgements}
The authors thank M. Douspis for helpful comments, P. Ferreira for
fruitful discussions and M. Santos for kindly providing his
code for the bispectrum in order to cross-check results. 
This work was partially 
supported by the CMBnet and the ACI-Jeunes Chercheurs 
``De la physique des hautes \'energies \`a la cosmologie observationnelle''. 
PGC is funded by the Funda\c{c}\~{a}o para a Ci\^{e}ncia e a Tecnologia,
MK acknowledges financial support by the Swiss National Science Foundation.
NA thanks TF and the University of Oxford for hospitality. 
PGC and MK further thank IAS Orsay for their hospitality.

\end{acknowledgements}


\end{document}